# CSMA-based Packet Broadcast in Radio Channels with Hidden Stations[1]

Authors: Yunpeng Zang (*Member IEEE*), Bernhard Walke (*Fellow IEEE*), Guido Hiertz (*Member IEEE*), Christian Wietfeld (*Senior Member IEEE*)

Abstract - Carrier Sense Multiple Access (CSMA) MAC protocols are known to suffer from the hidden station (HS) problem. The complete mathematical analysis of CSMA networks with HSs is still an open problem, even for broadcast communication with a simple linear network topology. In this paper we address this challenge by introducing a MAC layer modeling methodology based on time- and space-domain Markov processes. Using this methodology we derive the closed-form solution for the steady-state performance in infinite one-dimensional (1-D) CSMA networks with HSs. The analytical results are validated by simulation and establish that: 1) under the assumption of fixed frame duration, if the conditional channel access probability at each station exceeds a certain threshold, the CSMA system enters the status of "synchronized transmissions", where a large number of adjacent stations transmit in overlap and interfere each other resulting in null system goodput. 2) The maximum system goodput of CSMA broadcast communication increases with increased station density but becomes increasingly sensitive to the conditional channel access probability. In [25] we validate the analytical results gained in this paper by simulation of a multi-lane highway scenario, and provide quantitative guidance for congestion control algorithms in vehicular networks.

I. INTRODUCTION

Carrier Sense Multiple Access (CSMA) protocol for Medium Access Control (MAC) is based on the principle "listen before talk" that suffers from hidden stations (HSs) in multi-hop scenarios [18]. Though the HS problem has been intensively studied since it was discovered, a mathematical model with a closed-form solution for quantitative performance analysis of CSMA with HSs is still missing. This work addresses this challenge for CSMA broadcast communication in one-dimensional (1-D) linear network with an infinite number of stations. A practical example of CSMA broadcast communication with linear network topology is the IEEE 802.11p based vehicle-to-vehicle communication for vehicular safety services, as studied in [5].

---

[1] This work has been submitted to the IEEE for possible publication. Copyright may be transferred without notice, after which this version may no longer be accessible.

For steady-state analysis, we model any kind of CSMA protocol using the conditional channel access probability $p_{tx}$, which is defined as the probability that a station accesses the channel conditioned on the event that the channel is sensed idle for a given period of time $\sigma$ [s]. Similar to [20] and [2] [16] [21], probability $p_{tx}$ at each station is assumed to be stationary for a system in equilibrium.

At a given point in time distinct channel sensing results at stations that are spatially distant from each other cause the HS problem. Bearing this in mind, we give formal definitions of the HS condition, the HS problem, and HS scenarios according to the relation among stations established through the channel sensing operation. Given these definitions we can represent a HS scenario using a directed graph, namely the HS graph.

For a given HS graph, we observe that the states of stations in different subgraphs, which are the results from stations' channel sensing operation at a given time, are statistically mutually independent. This observation leads to the fundamental assumption of the conditional independence, as introduced in Section V.B, which enables the mathematical expression of interactions among stations that are either in mutual channel sensing range or hidden to each other.

The formal definitions of the HS condition, the HS problem, and HS scenarios, together with the assumption of the stationary conditional channel access probability and the conditional independence assumption form the basis of our methodology for modeling the HS problem in CSMA networks. Using the proposed methodology, we develop a HS model for the steady-state analysis in infinite 1-D CSMA broadcast networks. This HS model takes the conditional channel access probability $p_{tx}$, the frame duration $L$, and the number of one-side neighbors $R$ as input parameters and provides closed-form solutions for performance metrics, such as distribution of inter-transmitter distance $d_{TX}$, mean duration of the channel busy period $T_{RB}$ at a station, interference-free probability $p_{IF}$ of a reception burst, and system goodput $G$. To our best knowledge, this is the first work providing the above analytical results based on closed-form solutions considering HSs. In [25] we exploit the results presented in this paper: we show that when combined with a specific protocol model, e.g. the one derived from [1], our analytical model enables precise analysis of HSs in IEEE 802.11p broadcast networks. It is also shown in [25] that the infinite 1-D topology used in this work closely approximates the performance of IEEE 802.11p broadcast in a multiple-lane highway scenario.

The paper is organized as follows: We first introduce the system model in Section II. With the help of simulation results Section III presents the open problems of modeling HSs in CSMA broadcast networks. Section IV reviews the related work. Section V presents our modeling methodology covering formal definitions for HS condition, HS problem, HS scenarios, as well as two fundamental assumptions. In Section VI we develop the analytical HS model for infinite 1-D CSMA networks. In Section VII we validate the HS model by simulation studies. Section VIII presents further analytical results and discusses the impacts of frame duration, network density and conditional channel access probability on the performance of CSMA broadcast communication networks. Section IX gives concluding remarks.

## II. SYSTEM MODEL

A CSMA compliant station senses the channel as idle, when no station in its channel sensing range $r$ [m] is transmitting. Otherwise it senses the channel as busy, if at least one station in its channel sensing range, but not the station itself, is transmitting. At a given time, a station may either be sensing the channel (receiving a frame with or without co-channel interference in case of busy channel) or transmitting, but not both simultaneously [10]. In this work, we assume that all stations follow the same CSMA protocol for broadcast and have identical receiver performance and transmit power.

As assumed in [10] for the slotted version of $p$-persistent CSMA, the time axis is slotted with slot size $\sigma$ [s]. Stations are assumed time-synchronized and to start transmission only at the beginning of a slot. Unless otherwise stated, $\sigma$ is used as the unit of time in the following.

In this study, all frames are of constant duration $l$ [s]. Given the assumption of slotted time axis, the frame duration can be expressed in the unit of time slot $\sigma$:

$$L = \left\lceil \frac{l}{\sigma} \right\rceil \quad (1)$$

where $\lceil \alpha \rceil$ takes the next integer equal to or greater than $\alpha$.

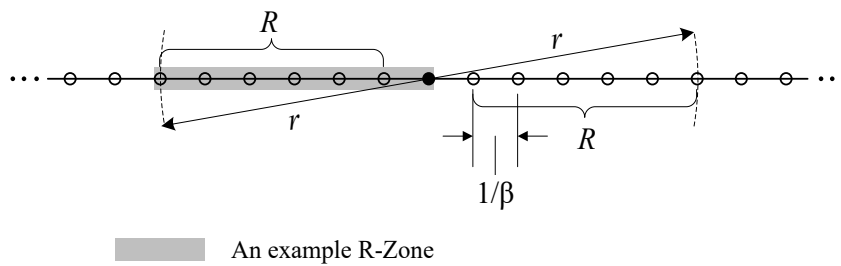

Figure 1: Topology of infinite 1-D network with HSs

Similar to [5], we concentrate on the 1-D network topology with an infinite number of stations, which are uniformly distributed with fixed spacing $1/\beta$ [m], where $\beta$ [stations/m] is the spatial density of stations. In this linear network, the number of stations $R$ in the one-side channel sensing range $r$ [m] is identical for all stations:

$$R = \lfloor r \cdot \beta \rfloor \qquad (2)$$

where $\lfloor \alpha \rfloor$ takes the last integer that is less than or equal to $\alpha$.

Figure 1 shows the studied network topology. With this network topology we define the distance measured in units of stations as *topological distance*. As shown in [25], the infinite 1-D topology used in this work closely approximates the performance of a multi-lane highway scenario.

In this infinite 1-D network, *R-Zone* is defined as the minimum spatial area covering $R + 1$ spatially contiguous stations. One property of R-Zone is that any transmission in a given R-Zone causes all other stations in the same R-Zone failing in sensing the channel idle. According to the listen-before-talk principle of CSMA, at a given point in time, if more than one station in a given R-Zone is transmitting, their transmissions must have started at the same time slot. According to our system model, each station belongs to $R + 1$ overlapping R-Zones.

### III. PROBLEM STATEMENT

Figure 2 shows simulation results of (a) the experimental Probability Mass Function (PMF) of inter-transmitter distance $d_{TX}$ of CSMA broadcast in a 1-D network and (b) the mean duration of channel busy period $T_{RB}$ depending on the conditional channel access probability $p_{tx}$.

The simulation scenario is shown in Figure 1, where all stations are placed in a linear topology with constant spacing. This simulation scenario avoids the border effect of any open-end topology by applying the periodic boundary condition. To emulate the infinite 1-D topology, we set the channel sensing range to be far less than the total length of the scenario. Table 1 gives the simulation parameters for generating results in Figure 2. The simulation code is available through [24].

Table 1. Simulation settings for generating results in Figure 2

| Parameters | Value |
|---|---|
| Total number of stations ($N$) | 800 |
| Transmission probabilities ($p_{tx}$) | 0.001 ~ 0.9 |
| Frame duration ($L$) | 16 $\sigma$, 32 $\sigma$ |
| Number of one-side neighbors ($R$) | 8, 16 |

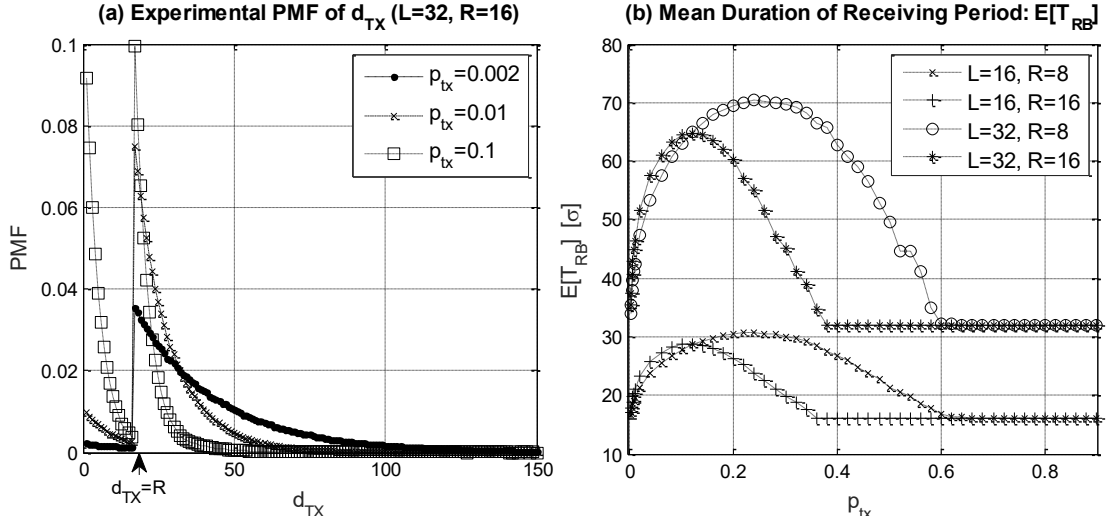

Figure 2: Simulation results showing the impacts of HSs in 1-D linear network

$d_{TX}$ is defined as the topological distance between two adjacent transmitting stations, say $TX1$ and $TX2$, at a given time. Figure 2 (a) shows that the experimental PMF of $d_{TX}$ in a 1-D network has a "double-peak" shape with peaks at $d_{TX} = 1$ and $d_{TX} = R + 1$, respectively. A similar "double-peak"-shaped PMF curve of $d_{TX}$ is reported in [15], where simulation studies with a more realistic channel model in multi-lane highway scenarios are performed.

$T_{RB}$ is the duration from the time that a station starts sensing the channel as busy to the time that the station senses the channel as idle again. As the simulation results in Figure 2 (b) show, for given values of $L$ and $R$ the mean value of $T_{RB}$ is greater than or equal to the frame duration $L$ depending on the value of $p_{tx}$. Overlapped asynchronous transmissions from HSs are the reason for the prolonged $T_{RB}$ value compared to $L$. According to Figure 2 (b), with an increased value of $p_{tx}$ the value of $T_{RB}$ converges back towards the frame duration $L$ after it has passed the maximum value. This is because assuming a fixed frame duration $L$, a higher $p_{tx}$ results in a higher probability of simultaneous transmission with $d_{TX} \leq R$, which can also be observed by comparing the PMF curves of $d_{TX}$ in Figure 2 (a) for different values of $p_{tx}$. The overlapped simultaneous transmissions result in high interference level at receivers and lead to null system goodput, when the value of $p_{tx}$ becomes high. In this study, this problem is referred to as "synchronized transmissions" under CSMA protocol.

An accurate analytical model for HS problem in CSMA networks shall be able to explain and model the above two observations, i.e. distribution of inter-transmitter distance $d_{TX}$ and prolonged duration $T_{RB}$ of channel busy period. To our best knowledge, this is the first work that gives closed-form solutions to both observations in infinite 1-D network provided parameters $L$, $R$ and $p_{tx}$.

IV.    RELATED WORK

HSs in CSMA networks are first addressed in [18]. However, an accurate analysis of the impact of HSs on the performance of a random access protocol is still an open issue [19], and finding the overlapping transmission times of HSs, as discussed in section III for the infinite 1-D network, is a serious challenge [11].

[8] [20] [12] and [19] model the HS problem in a CSMA network with limited number of hops, whereas [7] models the HS problem in looped-multi-hop 1-D and infinite 2-D ad-hoc CSMA networks. When calculating the impacts of hidden stations, [8] [20] and [7] assume constant number of stations contending for channel access and constant number of potential HSs or the number of HSs is calculated from deterministic spatial areas. The same assumptions are made in [13] and [22] for developing analytical models for IEEE 802.11p vehicular networks with an infinite 1-D topology taking account of the HS problem. As we show in Section III, these assumptions are not accurate in infinite 1-D network because both the number of stations contending for channel access and the number of potential HSs are random variables. [12] and [19] study the HS problem without considering the impacts from stations located two hops away. Therefore, approaches developed in [12] and [19] are not directly applicable to the infinite 1-D network studied in this work. Besides, none of the above papers give a quantitative analysis of the prolonged channel busy period $T_{RB}$ duration that is due to HSs. Authors of [5] model the broadcast performance of IEEE 802.11p in an infinite 1-D network. For simplicity reason, these authors assume HSs transmit independently according to a Poisson process. As shown in [23], this is a good approximation only when the conditional channel access probability $p_{TX}$ at each station is low. Our work solves the problems mentioned using joint time-domain and space-domain Markov chains.

Authors of [14] report that the HS problem is not an important issue in vehicular environments, whereas [15] shows the CSMA based IEEE 802.11p protocol suffers from the congestion problem in dense stations scenarios involving HSs. However, both [14] and [15] are based on simulation studies and both do not provide a quantitative analysis of the CSMA protocol in HS scenarios. This work provides the closed-form solution revealing the relation between the performance of CSMA in an infinite 1-D network and three fundamental parameters: frame duration $L$, single-side number of neighboring stations $R$ and conditional channel access probability $p_{tx}$.

## V. CSMA PROTOCOL AND HIDDEN STATION (HS) PROBLEM

### A. Conditional Channel Access Probability of CSMA Protocol

The HS model developed here is applicable to any CSMA protocol, as far as the steady state behavior of a station following the protocol can be described by the following paragraph:

When a station has sensed the channel as idle for a given duration, i.e. a slot time ($\sigma$[s]), with probability $p_{tx}$ the station starts to transmit and with probability $1 - p_{tx}$ the station does not transmit. If the channel is sensed busy, i.e. at least one station in its channel sensing range is transmitting, the station waits for the channel to return to idle state [10].

$p_{tx}$ is the channel access probability of a station conditioned on the channel being idle for a slot time $\sigma$ [s] in a CSMA system in equilibrium. $p_{tx}$ is different from any protocol-specified transmission probability, e.g. the parameter $p$ in $p-$ persistent CSMA protocol [10], which is a configurable parameter conditioned on the channel is sensed idle for $\sigma$ [s] and at least one frame is ready for transmission, whereas $p_{tx}$ is only conditioned on that the channel is sensed idle for $\sigma$ [s]. Using $p_{tx}$ instead of a protocol-specified transmission probability is essential for developing a model that is applicable to arbitrary CSMA protocols, e.g. IEEE 802.11p as shown in [25].

For studying the steady-state performance of CSMA protocols, we make the following assumption about the stationary conditional channel access probability:

**Assumption 1:** For a system in equilibrium, when a station senses the channel as idle for a given duration $\sigma$, the conditional channel access probability $p_{tx}$ that the station starts to transmit is independent of the temporal point that this probability is measured.

### B. Hidden Station and Hidden Station Problem

For a station $x$ in a given scenario, the *Clear-Set* $C(x)$ of $x$ is defined as the set of stations that satisfy

$$Pr\{the\ station\ is\ transmitting\ |\ station\ x\ senses\ the\ channel\ as\ idle\} = 0 \qquad (3)$$

Obviously, station $x$ itself is an element of set $C(x)$.

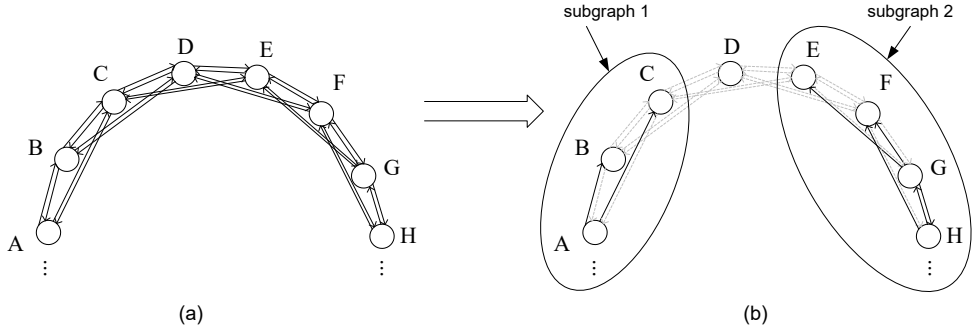

Figure 3: (a) HS graph of infinite 1-D network (b) subgraphs resulting from clear

If stations $x$, $y$ and $z$, where $x \neq y$, satisfy

$$Pr\{x \in C(z) | x \in C(y), z \in C(y)\} = 0 \qquad (4)$$

$x$ is defined to be *hidden* to $z$ with respect to $y$. At a given time, the *HS problem* occurs at station $y$, if $x$ and $z$ transmit simultaneously. It has to be noted that for the criterion in (4), it is not necessary that $z$ is hidden to $x$ with respect to $y$.

The HS problem is a receiver side problem and has the same effect at all stations having the same Clear-Set. In this article we focus on the infinite 1-D network topology described in section II, where each station has a unique Clear-Set different from others. We use directed graphs [4], which are referred to as *HS graphs* in this work, to represent the topology of HS scenarios. Figure 3 (a) shows the HS graph of an infinite 1-D network. Vertices in the HS graph exemplify stations and edges exemplify pairs of stations, where one station is in the Clear-Set of others. The direction of the edge is pointing from $x$ to $y$, if

$$x \in C(y) \text{ AND } x \neq y \qquad (5)$$

We use $D(V, E)$ to denote the directed graph, where $V$ is the vertex set and $E$ is the edge set of graph $D$. If a HS graph consists of at least one vertex that is not isolated, i.e. having at least one neighbor (see §1.2 in [4]), we call the scenario represented by this HS graph a *HS scenario*.

From the HS graph of the infinite 1-D topology, we observe that when one station, e.g. $D$ in Figure 3 (a), senses the channel idle, the graph can be updated by deleting edges originating from this station as well as edges originating from all stations in its Clear-Set, i.e. $B$, $C$, $E$ and $F$ in Figure 3 (a). This update is referred to as *clear channel operation* on a HS graph. Clear channel operation results in disjoint subgraphs, e.g. subgraph 1 and subgraph 2 in Figure 3 (b).

Besides, we define the event that a station senses the channel as busy or as idle or transmits a frame as the *state* of the station.

For a HS graph, we make the following conditional independence (CI) assumption:

**Assumption 2**: If a clear channel operation on a HS graph results in more than one disjoint subgraphs, the state of stations in one of the resulting subgraphs is independent from the state of stations in other resulting subgraph(s).

About clear channel operation it is worth noting that according to CSMA protocol, a station always attempts to access the channel with probability $p_{tx}$ after it senses the channel as idle for a slot time. Therefore, in this study the channel is not considered as idle if a station refrains from accessing the channel even if no signal is detected at that moment, e.g. due to Inter-Frame Spaces or the Network Allocation Vector in IEEE 802.11 [9].

More generic definitions of HS graph and clear-channel operation involving in multiple stations having identical Clear-Sets and on more generic network topologies than the infinite 1-D network can be found in [23] for studying the HS problem in more general CSMA networks.

## VI. HS MODEL FOR CSMA BROADCAST IN INFINITE 1-D NETWORK

The challenges of modeling HS problem and the prolonged duration of the channel busy period $T_{RB}$ come from the infinite number of ways (each with non-zero probability) that transmissions from HSs, with respect to a receiving station, may overlap with each other. We solve this problem for infinite 1-D network from a joint time- and space-domain perspective. Our approach includes three steps: In section VI.A we build a time-domain Markov chain describing the state of a certain station along the time axis at a certain spatial location; in section VI.B we build a space-domain Markov chain describing the state of stations along the spatial axis at a given point in time; at the end, the solution is found based on the ergodicity ([6] p.270) assumption about the infinite 1-D system: The overall probability that a station senses the channel as idle over the time shall be equal to the overall probability of a station is located in the interference-free spatial area throughout the scenario at a given point in time.

The HS model is developed based on Assumption 1 and Assumption 2 given in section V. The developed HS model takes three parameters: the conditional channel access probability $p_{tx}$, the frame duration $L$, and the number of stations $R$ in one-side channel sensing range. Performance metrics of the protocol such as mean duration of channel idle or channel busy period, interference-free probability of a reception burst and system goodput of CSMA broadcast are derived using the developed time- and space-domain Markov chains.

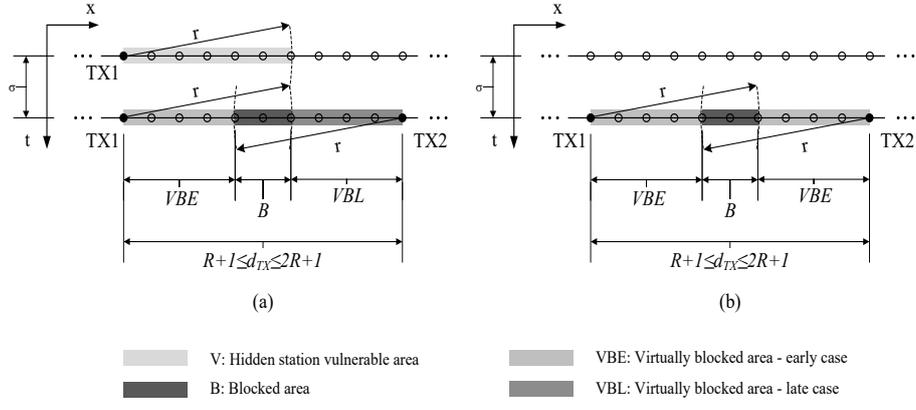

| V: Hidden station vulnerable area | VBE: Virtually blocked area - early case |
| B: Blocked area | VBL: Virtually blocked area - late case |

Figure 4: State of stations when $R+1 \leq d_{TX} \leq 2R+1$. (a) *TX1* and *TX2* start transmit simultaneously. (b) *TX1* starts transmit earlier than *TX2*

## A. Per-Station Time-Domain Markov Chain

### A.1 State of Station in HS Scenario

Without loss of generality, we analyze the state of a station between two adjacent transmitting stations, say $TX1$ and $TX2$, at any given time. $d_{TX}$ is the topological distance between $TX1$ and $TX2$. Depending on the value of $d_{TX}$ a station between $TX1$ and $TX2$ is in one of the following six *station states* at a given time:

*Idle (I)*: the station senses the channel as idle, i.e. none of $TX1$ and $TX2$ is in the channel sensing range of the station. This state requires $d_{TX} \geq 2R + 2$, as shown in Figure 5(a).

*HS Vulnerable (V)*: the station senses the channel as busy and the received signal comes from either $TX1$ or $TX2$, but not from both simultaneously. Besides, stations in this state are vulnerable to the HS problem, e.g. in Figure 5(a) when a station in the area *I* starts a transmission in the next time slot may result in the HS problem at some stations in the area *V*. This state also requires $d_{TX} \geq 2R + 2$.

*Blocked (B)*: the station is between $TX1$ and $TX2$ and senses the channel as busy because it receives signals from both $TX1$ and $TX2$ simultaneously. This state may happen when $d_{TX} \leq R$ and $R + 1 \leq d_{TX} \leq 2R + 1$, as shown in Figure 5(a) and in Figure 4, respectively.

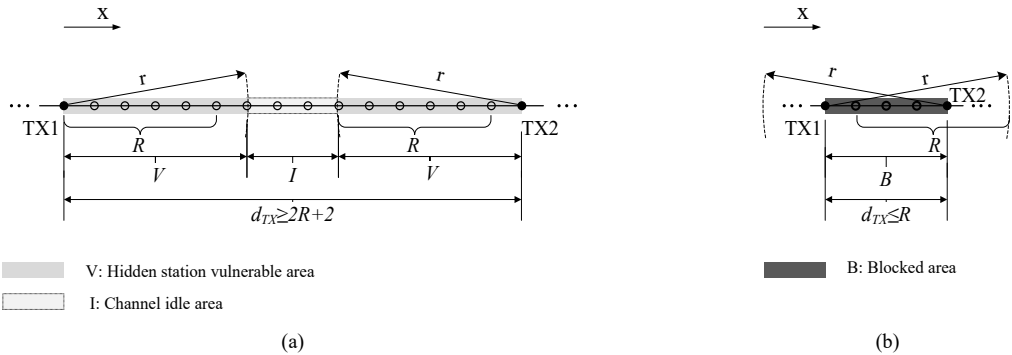

Figure 5: State of stations between *TX1* and *TX2*: (a) $d_{TX} \geq 2R+2$; (b) $d_{TX} \leq R$

*Virtually Blocked-Early Case (VBE)* and *Virtually Blocked-Late Case (VBL)*: the station is virtually blocked, if it senses the channel as busy and receives the signal from either $TX1$ or $TX2$, but not from both simultaneously, and it is not vulnerable to the HS problem before any transmission from $TX1$ or $TX2$ finishes. This state happens only if $R + 1 \leq d_{TX} \leq 2R + 1$, as shown in Figure 4. We further differentiate the virtually blocked states into two cases: as shown in Figure 4(a), if the station is receiving the signal from the later transmitter of $TX1$ and $TX2$, it is in $VBL$ state, otherwise it is in $VBE$ state. Note, the $VBE$ case also covers the situation that TX1 and TX2 start transmissions simultaneously, as shown in Figure 4 (b).

*Transmitting (TX)*: The station is transmitting a frame at the given time, e.g. the station $TX1$ or $TX2$ in Figure 5.

Besides, we define a super-state *Receiving Busy (RB)* comprising states $V, B, VBE$ and $VBL$.

When we talk about the spatial occupancy of stations, $I, V, VBE, VBL$ and $B$ are also used to denote the maximal spatial areas composed of contiguously located stations that are in the same denoted station state, e.g. as already shown in Figure 5 and Figure 4.

## A.2  Per-Station Time-Domain Markov Chain

The per-station time-domain Markov chain for CSMA broadcast communication is shown in Figure 6. There, transitions take place at the start of each time slot $\sigma$ and all states in the Markov chain have concrete meaning in time. For the same reason, station states introduced in the previous section consist of sequential Markov chain states, which are referred to as sub-states of that station state. The number of sub-states in a station state represents the time span in the unit of time slot $\sigma$ of that station state. Station state $I$ always has one sub-state. States with dashed circles in Figure 6 are duplicated symbols for Markov chain states with the same name for the sake of easy presentation.

$I$ is the state that a station senses the channel as idle in the current time slot.

$TX_{(L,n)}$, $n = 1,2, ..., L$, are sub-states of $TX$. The first index value is the total duration $L$ of $TX$, i.e. the frame length, counted in the unit of $\sigma$. The second index $n$ indicates the current time slot to be the $n$-th time slot of the frame in transmission. As a convention for denoting a sub-state in the time-domain Markov chain, subscript $(l, n)$ always denotes the $n$-th sub-state of a state with a total length $l$, where $1 \leq l \leq L$ and $1 \leq n \leq l$.

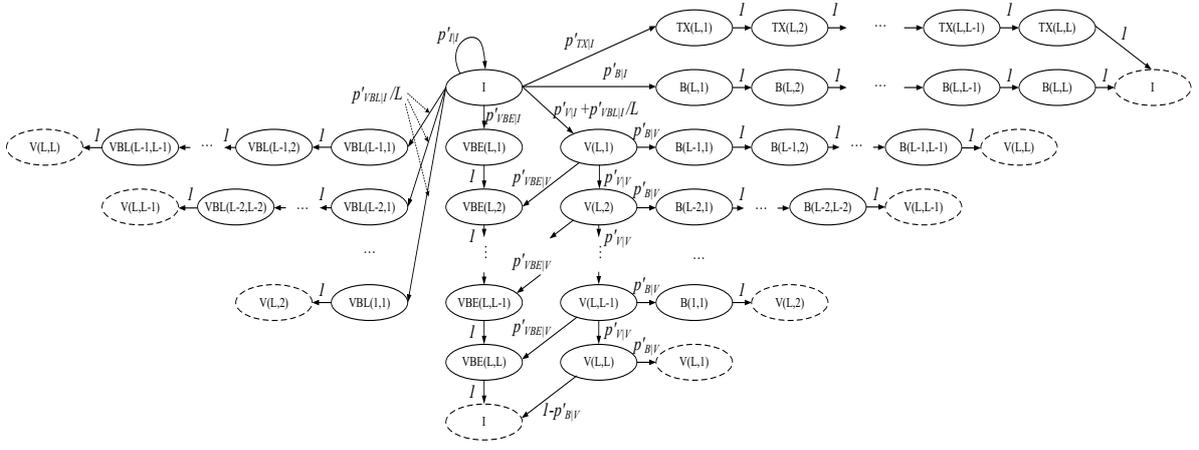

Figure 6: Time-domain Markov chain for CSMA broadcast in infinite 1-D network

$B_{(L-m,n)}$, $m = 0,1, \ldots, L-1$; $n = 1,2, \ldots, L-m$, are sub-states of $B$. As a station may enter station state $B$ from a sub-state of $V$, $B$ may have a total length less than $L$, e.g. $B_{(L-m,n)}$, if entered from sub-state $V_{(L,m)}$.

$V_{(L,n)}$ and $VBE_{(L,n)}$, $n = 1,2, \ldots, L$ are sub-states of $V$ and $VBE$, respectively.

$VBL_{(L-m,n)}$, $m = 1,2, \ldots, L-1$; $n = 1,2, \ldots, L-m$, are sub-states of $VBL$. The maximum total length of state $VBL$ is $L - 1$ because a station in this state is synchronized to the later transmitter, which implies that the earlier transmitter has been transmitting for at least one time slot. A station enters $VBL$ of a total length $L - m$, when the early transmitter has been transmitting for $m$ time slots.

Before discussing the time-domain Markov chain and its state transition probabilities, we first introduce the following nine supporting probabilities:

$p'_{I|I}$, $p'_{TX|I}$, $p'_{V|I}$, $p'_{B|I}$, and $p'_{VBE|I}$, are the probabilities that a station stays in state $I$, starts transmission, enters station states $V$, $B$, or $VBE$ in the next time slot conditioned on it senses the channel as idle in the current time slot, respectively.

$p'_{VBL|I}$ is the probability that a station enters $VBL$ in the next time slot, conditioned on it senses the channel as idle in the current time slot, as depicted in Figure 4 (a), or a station falls in the area $a$ depicted in Figure 7 in the next time slot, conditioned on the station senses the channel as idle in the current time slot, if the late transmitter starts to transmit right after the early transmitter finishes its transmission.

$p'_{B|V}$ is the probability that a station enters $B$ in the next time slot, conditioned on it is in $V$ in the current time slot, as depicted in Figure 4 (a), or a station falls in the area $b$ depicted in Figure 8 in the next time slot conditoned on it is in $V$ in the current time slot, if the late transmitter starts to transmit right after the early transmitter finishes its transmission.

$p'_{VBE|V}$ is the probability that a station enters $VBE$ in the next time slot conditioned on it is in $V$ in the current time slot, as shown in Figure 4 (a).

$p'_{V|V}$ is the probability that a station stays in $V$ in the next time slot conditioned on it is in $V$ in the current time slot.

Let $s(t)$ denotes the state of a station at time $t$. Given arbitrary state $A$ and state $B$ in a Markov chain, for the one step transition probability from $A$ at time $t$ to $B$ at time $(t+1)$, we use the short notation $P\{B_{t+1}|A_t\} = P\{s(t+1) = B|s(t) = A\}$ [17]. Here time $t$ is normalized to a slot time $\sigma$. Let $\pi_A = \lim_{t \to \infty} \Pr\{s(t) = A\}$ be the limiting probability of state $A$ in the Markov chain. $\boldsymbol{\pi} = (\pi_1, \pi_2, \dots, \pi_n)$ is the state distribution vector of the Markov chain, with $n$ being the total number of states. The time independent one-step state transition probability from state $A$ to state $B$ is denoted as $p_{B,A}$ [17].

All non-zero transition probabilities in the time-domain Markov chain are given in Figure 6 and expressed in (6). About the transition probabilities in the time-domain Markov chain, the following is worth noting:

Once a station enters state $TX_{(L,1)}$, it transits towards state $TX_{(L,L)}$ with probability 1.

Once a station enters any sub-state of $B$, $VBE$, or $VBL$ it has to wait for one or both of the transmitters in its channel sensing range finishing the transmission. This is the reason for transition probabilities of 1 among sub-states of $B$, $VBE$, or $VBL$. When one or both transmitters finish the transmission, the station transits to the corresponding sub-states of $V$, if the last state is any of $B_{(L-m,L-m)}$ and $VBL_{(L-m,L-m)}$; $m = 1,2,...L-1$, or to state $I$, if the last state is $VBE_{(L,L)}$.

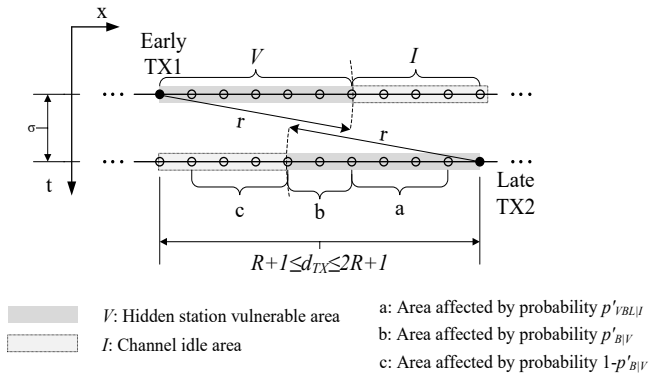

Figure 7: Special case where the early transmitter stops when the late transmitter start to transmit

$$\begin{cases} p_{I,I} = p'_{I|I} \\ p_{TX_{(L,1)},I} = p'_{TX|I} \\ p_{B_{(L,1)},I} = p'_{B|I} \\ p_{V_{(L,1)},I} = p'_{V|I} + \dfrac{p'_{VBL|I}}{L} \\ p_{VBE_{(L,1)},I} = p'_{VBE|I} \\ p_{VBL_{(L-m,1)},I} = \dfrac{p'_{VBL|I}}{L} & ,m = 1,2,...,L-1 \\ p_{TX_{(L,n+1)},TX_{(L,n)}} = 1 & ,n = 1,2,...,L-1 \\ p_{I,TX_{(L,L)}} = 1 \\ p_{B_{(L-m,n+1)},B_{(L-m,n)}} = 1 & ,m = 0,1,...,L-2 \\ & ,n = 1,2,...,L-m-1 \\ p_{V_{(L,L-m+1)},B_{(L-m,L-m)}} = 1 & ,m = 0,1,...,L-1 \\ p_{B_{(L-n,1)},V_{(L,n)}} = p'_{B|V} & ,n = 1,2,...,L-1 \\ p_{VBE_{(L,n+1)},V_{(L,n)}} = p'_{VBE|V} & ,n = 1,2,...,L-1 \\ p_{V_{(L,n+1)},V_{(L,n)}} = p'_{V|V} & ,n = 1,2,...,L-1 \\ p_{V_{(L,1)},V_{(L,L)}} = p'_{B|V} \\ p_{I,V_{(L,L)}} = 1 - p'_{B|V} \\ p_{VBE_{(L,n+1)},VBE_{(L,n)}} = 1 & ,n = 1,2,...,L-1 \\ p_{I,VBE_{(L,L)}} = 1 \\ p_{VBL_{(L-m,n+1)},VBL_{(L-m,n)}} = 1 & ,m = 1,2,...,L-2 \\ & ,n = 1,2,...,L-m-1 \\ p_{V_{(L,L-m+1)},VBL_{(L-m,L-m)}} = 1 & ,m = 1,2,...,L-1 \end{cases} \quad (6)$$

In every sub-state of $V$ there are non-zero probabilities that the station transits to other states, e.g. to states $VBE$ or $B$, because any station in the neighboring $I$ area affecting this station may start to transmit in the next time slot. More specifically, a station in sub-state $V_{(L,n)}$, $n = 1,2,...L-1$ may transit to $B_{(L-n,1)}$ with probability $p'_{B|V}$, or to $VBE_{(L,n+1)}$ with probability $p'_{VBE|V}$, as shown in Figure 4 (a). Otherwise, the station stays in state $V$ with probability $p'_{V|V} = 1 - p'_{B|V} - p'_{VBE|V}$. A special case is sub-state $V_{(L,L)}$, i.e. the last sub-state of $V$. When a HS starts to transmit in the next time slot, instead of transiting to $B$ or $VBE$, the station transits to sub-state $V_{(L,1)}$ with probability $p'_{B|V}$ if it is located in area $b$ in Figure 7, or the station transits to state $I$ with probability $(1 - p'_{B|V})$, if it is located in area $c$ in Figure 7. Owing to Assumption 1 and Assumption 2, transition probabilities are stationary in time and identical to all sub-states of $V$.

According to Assumption 1 and Assumption 2, the starting times of transmissions from stations $A$ and $B$ are independent, if $A$ and $B$ do not belong to a same R-Zone and no other station between $A$ and $B$ is in state $TX$. This is the reason for calculating the transition probability $P_{VBL_{(L-m,1)},I}$ as $\frac{p'_{VBL|I}}{L}$ for $1 \leq m \leq L-1$ in (6).

The Markov chain in Figure 6 models a CSMA protocol that requires stations to sense the channel idle before sending any frame, i.e. no consecutive packet transmission without channel sensing is allowed at a station. This explains why transition probabilities from $TX_{(L,L)}$ to $I$, from $B_{(L,L)}$ to $I$, and from $VBE_{(L,L)}$ to $I$ are modeled equal to 1.

For the infinite 1-D network, the nine supporting probabilities contained in (6) are determined by the conditional channel access probability $p_{tx}$, the frame duration $L$ and the distribution of the *interference free area* $d_F$. It is observed that in the infinite 1-D scenario and under Assumption 2 the size of interference free area $d_F$ measured in topological distance, at any given time, follows a geometric distribution, provided that all stations access the channel with probability $p_{tx}$. Based on this observation, all nine supporting probabilities introduced in this section can be expressed using parameter $p_{O,F}$ of the geometry distribution of $d_F$ together with other known parameters, i.e., $p_{tx}$, $L$, and $R$. Calculation of all supporting probabilities would exceed page limits of this paper and is available in Appendix B of [23][2].

## B. Space-Domain Markov Chain for Inter-Transmitter Distance

The space-domain Markov chain is constructed by traversing the 1-D scenario at a given time. Without loss of generality, we take the traversing direction from left to right, as indicated in Figure 9. We study area $O$, which is occupied by signal(s) from transmitting stations, and the interference-free area $F$, where stations sense the channel as idle at this given time. See Figure 9 for an example of area $O$ containing four transmitting stations and the interference-free areas $F$ next to area $O$.

Figure 8 shows the space-domain Markov chain. Transitions in this Markov chain happen when one traverses from one station to the next station in direction x indicated in Figure 9. It is worth emphasizing that during transitions in the space-domain Markov chain the time is frozen at a given time point.

States in the Markov chain describe the status of channel at the current spatial location. The state can be either $F$ for free or $O$ for occupied.

---
[2] Appendix B of [23] is supplied as Annex D to this manuscript.

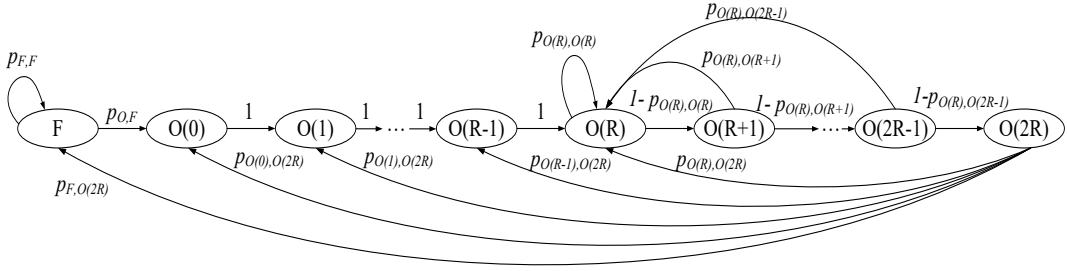

Figure 8:   Space-domain Markov chain

Analogously to the division of station state $TX$ into sub-states in the time-domain Markov chain, we divide the $O$ state in the space-domain Markov chain into sub-states $O(n)$, $n = 0, 1, \ldots, 2R$. If the station at the current location is transmitting, we define the system is in sub-state $O(R)$. Sub-states $O(n), n = 0, 1, \ldots, R-1$, mean that the station at the current location is not transmitting and one has traversed more than $R$ steps since the last transmitting station and one needs to traverse $R - n$ more steps to reach the next location of a transmitting station. Sub-states $O(n), n = R+1, R+2, \ldots, 2R$, mean that the station at the current location is not transmitting and one has traversed $n - R$ steps since the last transmitting location. $O(2R)$ is the only sub-state where the system may either leave the occupied area to enter a free area or continue to stay in the occupied area for at least $R + 1$ more steps, which corresponds to the transition probability $p_{O(R),O(2R)}$, as shown in Figure 8. The space-domain Markov chain describes all possible occupied areas and interference-free areas in the infinite 1-D network at a given time.

From Assumption 2, it is deduced that the number of steps that the system can stay in area $F$, i.e. the size $d_F$ of an interference free area $F$ measured in topological distance, follows a geometric distribution with parameter $p_{O,F}$. $p_{O,F}$ is the probability that the system is in the occupied area in the next step, conditioned on that the system is in the interference-free area in the current step. The Probability Mass Function (PMF) of $d_F$ is

$$f_{d_F}(k) = Pr\{d_F = k\} = (1 - p_{O,F})^{k-1} \cdot p_{O,F}; \quad k = 1, 2, \ldots \quad (7)$$

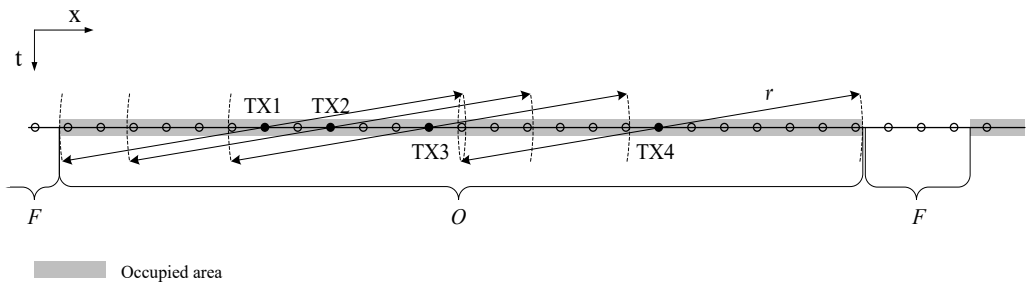

Figure 9:   Occupied area $O$ and free areas $F$ in space domain at a given point in time

Transition Probabilities of Space-domain Markov Chain

All non-zero transition probabilities in the space-domain Markov chain shown in Figure 8 can be calculated based on the distribution of $d_F$ and other known parameters, i.e. $p_{tx}$, $L$, and $R$.

$p_{F,F}$ is the probability that the next location is also in an interference-free area, conditioned on the current location is in a interference-free area: $p_{F,F} = 1 - p_{O,F}$.

$p_{O(n+1),O(n)}$, $n = 0, 1, \ldots, R - 1$, is the transition probability among non-transmitting sub-states in the occupied area at the left side of the next transmitting location, as shown in Figure 9. As each sub-state has a deterministic topological distance to the next transmitting location, this transition probability equals 1.

$p_{O(R),O(n)}$, $n = R, R + 1, \ldots, 2R - 1$, is the probability that the next location is a transmitting location conditioned on that one has traversed $n - R$ steps since the last transmitting location. It can be calculated using its relation to the PMF of inter-transmitter topological distance $d_{TX}$:

$$Pr\{d_{TX} = d\} = \begin{cases} p_{O(R),O(R)} & , d = 1 \\ \prod_{j=R}^{R+d-2} (1 - p_{O(R),O(j)}) \cdot p_{O(R),O(R+d-1)} & , 2 \leq d \leq R \end{cases} \quad (8)$$

The PMF of $d_{TX}$ for $1 \leq d_{TX} \leq R$ is given later in (11).

$p_{O(n+1),O(n)}$, $n = R, R + 1, \ldots, 2R - 1$, is the probability that the next location is not a transmitting location conditioned on that one has traversed $n - R$ steps since the last transmitting location. Based on the space-domain Markov chain, we have $p_{O(n+1),O(n)} = 1 - p_{O(R),O(n)}$.

$p_{O(n),O(2R)}$, $n = 0, 1, \ldots, R$, is equal to the probability that the inter-transmitter distance $d_{TX} = (R - n) + R + 1$, which is given by the PMF of $d_{TX}$ for $R + 1 \leq d_{TX} \leq 2R + 1$ in (12).

$p_{F,O(2R)}$ is the probability that $d_{TX} \geq 2R + 2$ given by the PMF of $d_{TX}$ in (13). This is also the probability that the next location is in an interference-free area, conditioned on the current location in an occupied area.

All transition probabilities in the space-domain Markov chain are expressed using known parameters $p_{tx}$, $L$, $R$ and the single unknown $p_{O,F}$, as summarized in (9):

$$\begin{cases} p_{F,F} = 1 - p_{O(0),F} \\ p_{O(0),F} = p_{O,F} \\ p_{O(n+1),O(n)} = 1 & ,n = 0,1,\dots,R-1 \\ p_{O(R),O(n)} = (1-p_{tx})^{n-R} \cdot p_{tx} \cdot (1-p_{O,F})^{n-R+1} & ,n = R, R+1,\dots,2R-1 \\ & ,n = R, R+1,\dots,2R-1 \\ p_{O(n),O(2R)} = \dfrac{L \cdot p_{tx} \cdot a^{R-n}}{1 + L \cdot p_{tx} \cdot \dfrac{1-a^{R+1}}{1-a}} & ,n = 0,1,\dots,R \\ p_{F,O(2R)} = \dfrac{1}{1 + L \cdot p_{tx} \cdot \dfrac{1-a^{R+1}}{1-a}} \end{cases} \quad (9)$$

where

$$a = (1 - p_{tx}) \cdot (1 - p_{O,F}) \tag{10}$$

The distribution of $d_{TX}$ in infinite 1-D work can be calculated using the geometric distribution of $d_F$ and based on Assumption 1 and Assumption 2. Due to the limited space, only the calculation results are summarized here. The detailed mathematical derivations are documented in Section 5.3.3 of [23][3].

$$f_{d_{TX}}(k) = Pr\{d_{TX} = k\} = (1-p_{tx})^{k-1} \cdot p_{tx} \cdot (1-p_{O,F})^k \quad ,1 \leq k \leq R \tag{11}$$

$$\begin{aligned} f_{d_{TX}}(k) &= \\ &= \left[1 - p_{tx} \cdot (1-p_{O,F}) \cdot \frac{1-a^R}{1-a}\right] \cdot \frac{L \cdot p_{tx} \cdot a^{k-(R+1)}}{1 + L \cdot p_{tx} \cdot \frac{1-a^{R+1}}{1-a}} \\ &,R+1 \leq k \leq 2R+1 \end{aligned} \tag{12}$$

$$\begin{aligned} f_{d_{TX}}(k) &= \\ &= \left[1 - p_{tx} \cdot (1-p_{O,F}) \cdot \frac{1-a^R}{1-a}\right] \cdot \frac{1}{1 + L \cdot p_{tx} \cdot \frac{1-a^{R+1}}{1-a}} \cdot (1-p_{O,F})^{k-2R-2} \cdot p_{O,F} \\ &,k \geq 2R+2 \end{aligned} \tag{13}$$

where $a$ is given in (10).

## C. Joint Solution for Time- and Space-Domain Markov Chains

(6) and (9) give all non-zero transition probabilities expressed by three known parameters: $p_{tx}$, $L$, and $R$, and an unknown parameter $p_{O,F}$, in the time- and space- domain Markov chains, respectively. To solve for $p_{O,F}$, we use the ergodicity assumption of the system in both time- and space- domains. This means the probability that a station senses the channel as idle in the time-domain shall be equal to the probability that a station is located in an interference-free area $F$ in the space-domain. Therefore, we have

$$\pi_I = \pi_F \tag{14}$$

---

[3]Section 5.3.3 of [23] is supplied as Annex A to this manuscript.

Both solutions of $\pi_I$ and $\pi_F$ rely on the unknown probability $p_{O,F}$. By enforcing (14) we get an equation system, where the solution for $p_{O,F}$ can be found using numerical methods. The limiting probability of states in the time-domain and the space-domain Markov chains are also solved using numerical methods.

### D. Performance Metrics

#### D.1 Time Metrics

Figure 10 illustrates time metrics of the CSMA broadcast protocol measured at a station in the infinite 1-D network. Mean values of these metrics are calculated using the limiting distributions of the time-domain Markov chain.

$T_I$ is the length of the *channel idle period*. As shown in Figure 6, state $I$ has the fixed self-transition probability $p'_{I|I}$. Thus, the duration that a station stays in state $I$ can be modeled using geometric distribution with parameter $(1 - p'_{I|I})$, and the expected length of $T_I$ is

$$\overline{T}_I = \sum_{n=1}^{\infty} n \cdot (p'_{I|I})^n \cdot (1 - p'_{I|I}) = \frac{1}{1 - p'_{I|I}} \tag{15}$$

$T_{NI}$ is the length of the *non-idle period* between two consecutive channel idle period. $T_{NI}$

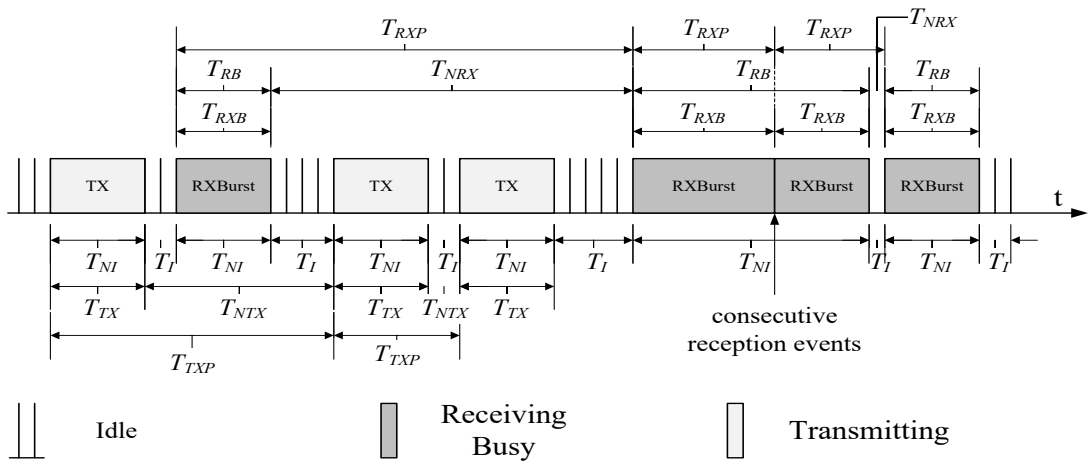

Figure 10: Time-domain metrics in HS model

may be either a transmission period $T_{TX}$ or a channel busy period $T_{RB}$. Here, we assume the occurrence of event "channel goes from non-idle to idle" is a renewal process [3], i.e., $T_I + T_{NI}$ is an i.i.d. random variable. Given the expected length of channel idle period $\overline{T_I}$ in (15) we can calculate the expected value of $T_{NI}$:

$$\overline{T_{NI}} = \overline{T_I} \cdot \frac{1-\pi_I}{\pi_I} \tag{16}$$

$T_{TX} = L$ is the duration of a frame transmission.

$T_{NTX}$ is the duration measured from the ending time of a frame transmission to the starting time of the next frame transmission at a station. Assuming the occurrence of event "the station starts a frame transmission" is a renewal process, the expected value of $T_{NTX}$ is calculated as:

$$\overline{T_{NTX}} = T_{TX} \cdot \left(\frac{1}{\sum_{n=1}^{L} \pi_{TX_{(L,n)}}} - 1\right) \tag{17}$$

$T_{TXP}$ is the length of a *transmission period*, which is defined as the time between the starting point of two consecutive frame transmissions. The expected value of $T_{TXP}$ is:

$$\overline{T_{TXP}} = T_{TX} + \overline{T_{NTX}} \tag{18}$$

where, $\overline{T_{NTX}}$ is solved in (17).

$T_{RB}$ is the length of a *channel busy period*, during which a station continuously senses the channel busy. It may consist of one or more consecutive reception bursts, as shown in Figure 10. In this scenario, $T_{NI}$ is either a $T_{TX}$ or a $T_{RB}$ following a $T_I$. Therefore, the expected length of $T_{NI}$ can be calculated as

$$\overline{T_{NI}} = \frac{1 - p'_{I|I} - p'_{TX|I}}{1 - p'_{I|I}} \cdot \overline{T_{RB}} + \frac{p'_{TX|I}}{1 - p'_{I|I}} \cdot \overline{T_{TX}} \tag{19}$$

where, $\frac{1-p'_{I|I}-p'_{TX|I}}{1-p'_{I|I}}$ is the probability that a $T_I$ is followed by a $T_{RB}$, whereas $\frac{p'_{TX|I}}{1-p'_{I|I}}$ is the probability that a $T_I$ is followed by a $T_{TX}$. From (19), the expected length of a channel busy period $T_{RB}$ is:

$$\overline{T_{RB}} = \frac{(1 - p'_{I|I}) \cdot \overline{T_{NI}} - p'_{TX|I} \cdot L}{1 - p'_{I|I} - p'_{TX|I}} \tag{20}$$

where, $\overline{T_{NI}}$ is given in (16).

Being marked as "RXBurst" in Figure 10, $T_{RXB}$ is the length of a *reception burst*, which is defined as the duration from the time a station starts a reception to the time when the station detects the end of the received frame or the end of all received overlapping frames. The length of a reception burst may be longer than frame duration $L$ due to the overlapped transmission from HSs. The introduction of period $T_{RXB}$ is due to the consecutive reception event, as shown in Figure 10. $T_{NRX}$ is the length of *non-reception period*, in which the station may either be sensing the channel as idle or transmitting a frame. $T_{RXP}$ is the length of the *reception period* defined as the duration between the starting time of two consecutive reception bursts.

Similarly to the calculation of $\overline{T_{RB}}$, we calculate the mean values of $T_{RXB}$, $T_{NRX}$, and $T_{RXP}$ using the limiting distribution of the time-domain Markov chain. [Section 5.5.3 of [23] gives the detailed calculation of each metric][4]. Among these time metrics $\overline{T_I}$, $\overline{T_{TXP}}$, $\overline{T_{RB}}$, and $\overline{T_{RXP}}$ are of particular interest to our study.

### D.2    Interference-Free Probability of Reception Burst

As a metric of the reliability of CSMA broadcast, we define the interference-free probability $p_{IF}$ of a reception burst as the probability that a frame is received without being interfered by any other transmission, conditioned the receiver starts a reception burst. $p_{IF}$ is also the ratio of reception burst free from interference out of all reception bursts at a receiver. Besides, we are also interested in the $p_{IF}$ performance with respect to a given topological reception distance $d_{RX}$ from the receiver to the transmitter, i.e. the conditional distribution of $d_{RX}$ given that a frame is received interference-free $f_{d_{RX}|IF}(k), 1 \leq k \leq R$. The calculation of $p_{IF}$ and $f_{d_{RX}|IF}(k)$ is based on the limiting distribution of the time domain Markov chain and the geometric distribution of $d_F$. [[23], Section 5.5.4][5] provides the detailed calculation and [[23], Eq.(5.48)]5 and [[23], Eq.(5.49)]5 give the results of $p_{IF}$ and $f_{d_{RX}|IF}(k)$, for $1 \leq k \leq R$, respectively.

### D.3    Goodput of CSMA Broadcast

Goodput ($G$) is the metric for the throughput efficiency of CSMA broadcast communication. In this study, $G$ is calculated as:

$$G = \frac{L \cdot p_{IF}}{\overline{T_{RXP}}} \quad (21)$$

---
[4]Section 5.5.3 of [23] is supplied as Annex B to this manuscript.
[5]Section 5.5.4 of [23] is supplied as Annex C to this manuscript.

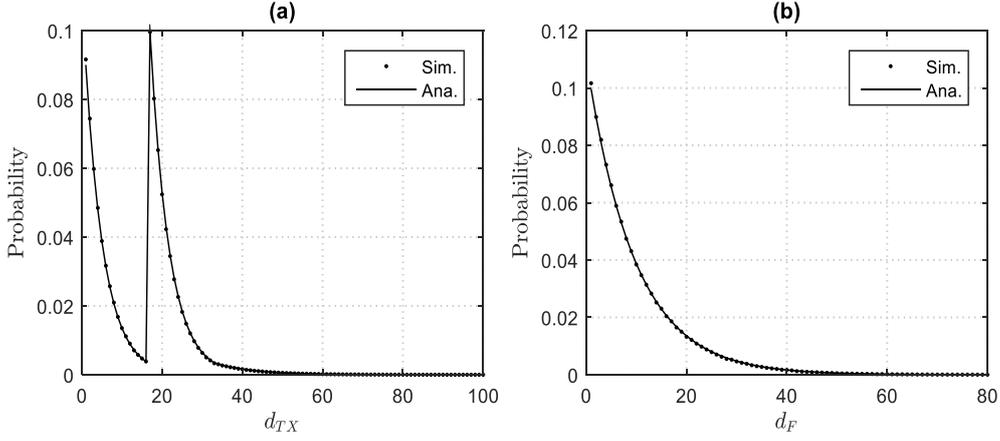

Figure 11:   Simulated and analytical PMF of $d_{TX}$ and $d_F$ for $L=32$, $R=16$ and $p_{tx}=0.1$,

Where, $\overline{T_{RXP}}$ is the expected value of the reception period discussed in Section VI.D.1 and $p_{IF}$ is the interference-free probability of a reception burst from Section VI.D.2. $G$ can also be understood as the fraction of time used for delivering interference-free frames out of the total time resource.

## VII.   MODEL VALIDATION

The HS model is validated by comparing the analytical results with Monte-Carlo simulation results using Matlab®. The simulation scenario and parameters are given in section III.

### A.1 Distribution of $d_{TX}$ and $d_F$

The analytical and simulated PMF results of inter-transmitter distance $d_{TX}$ (see (11) to (13)) and of the size of an interference-free area $d_F$ (see (7)) are shown in Figure 11 and Figure 12 for scenarios with $L = 32$, $R = 16$, and $p_{tx} = 0.1$ and $0.34$. The numerical solution for $p_{O,F}$, i.e. the parameter of the geometric distribution of $d_F$, is given in the caption of each figure.

In both figures, the simulated PMF of $d_{TX}$ matches the analytical solution for $f_{d_{TX}}(k)$ from (11) to (13) very well. The HS model successfully models the "double-peak"-shaped PMF curve of $d_{TX}$ with peaks at $d_{TX} = 1$ and $d_{TX} = R + 1$, as discussed in section III. This "double-peak"-shaped PMF curve of $d_{TX}$ is owing to the fact that transmitters having a distance less than or equal to $R$ are synchronized to each other, i.e. starting from the same time slot, whereas transmitters having a distance greater than $R$ may not be synchronized. In both ranges, $1 \leq d_{TX} \leq R$ and $R + 1 \leq d_{TX} \leq 2R + 1$, the shape of the curve is governed by (7).

The simulated results for $d_F$ in Figure 11 match the geometric distribution with parameter $p_{O,F} = 0.0996$ very well, while the simulated $d_F$ curve in Figure 12 statistically deviates from the analytically gained curve in some parts with a clear trend towards a geometric distribution. The geometric distribution of simulated $d_F$ results is a strong indication for the presence of the "memorylessness" property in the distribution of $d_F$, which can be taken as a confirmation of Assumption 2 in Section V.B.

By comparing the $d_{TX}$ curves in Figure 11 and Figure 12, one can see a much higher probability for $1 \leq d_{TX} \leq R$ at higher $p_{tx}$ values. Recall the fact that CSMA transmissions in a same R-Zone are always synchronized to each other. This high probability of $1 \leq d_{TX} \leq R$ indicates a problem of synchronized CSMA transmissions in large areas due to overlapped R-Zones in a HS scenario, when the value of $p_{tx}$ is high, in other words, when broadcast communication fails and goodput of stations is approaching zero, see Section VII.A.2.

*A.2 Station State Probability Distribution*

The limiting probability $\pi_I$ in the time-domain Markov chain is the probability that a station is sensing the channel as idle at any given time slot. The probability that a station is transmitting at a given slot is:

$$\pi_{TX} = \sum_{n=1}^{L} \pi_{TX_{(L,n)}} \qquad (22)$$

where $\pi_{TX_{(L,n)}}, 1 \leq n \leq L$, is the limiting distribution of state $TX_{(L,n)}, 1 \leq n \leq L$, in the time domain Markov chain. The probability that a station is in receiving busy state ($RB$) at any given time slot is:

$$\pi_{RB} = 1 - \pi_I - \pi_{TX} \qquad (23)$$

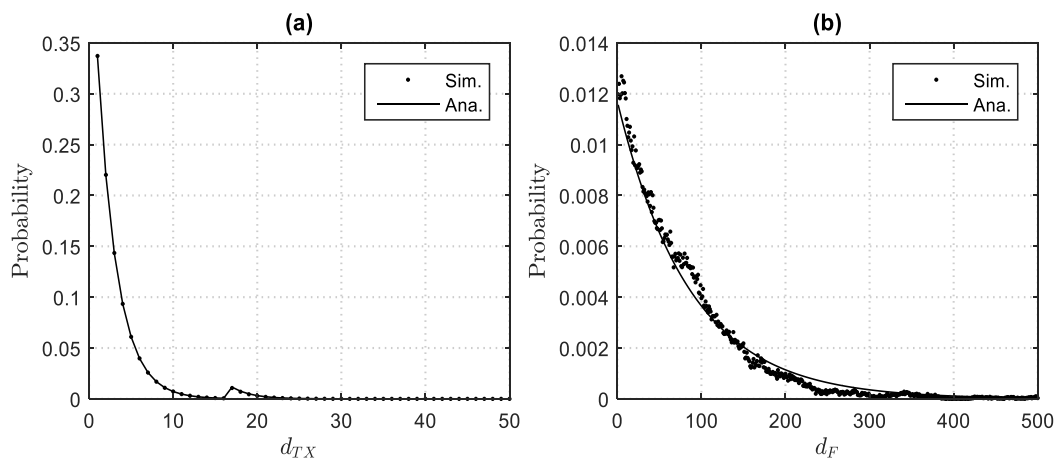

Figure 12: Simulated and analytical PMF of $d_{TX}$ and $d_F$ for *L=32, R=16* and *p$_{tx}$=0.34*, (solution: *p$_{OF}$=0.0116*)

The simulated and analytical steady state probabilities $\pi_I$, $\pi_{RB}$, and $\pi_{TX}$ against the conditional channel access probability $p_{tx}$ at a station are shown in Figure 13, where the two subplots are dedicated to scenarios with different $R$ values, as given in the title of each subplot. The well matched simulated and analytical results across the range of $p_{tx}$ values studied validate the analytical model and show its accuracy.

It is common to both scenarios that at the low end of $p_{tx}$, when $p_{tx}$ increases, $\pi_I$ drops rapidly, whereas the probability $\pi_{RB}$ rises very fast. Interestingly, when the value of $p_{tx}$ further increases, $\pi_I$ converges to the fixed value $1/(L+1)$ and $\pi_{RB}$ turns over from increasing to decreasing. At the high end of $p_{tx}$, there is a linear relation of both $\pi_{RB}$ and $\pi_{TX}$ dependent on $p_{TX}$, while $\pi_I$ stays constant at $1/(L+1)$. This is exactly the phenomenon of "synchronized" CSMA transmissions in HS scenarios when $p_{tx}$ becomes high, as discovered in the last section. After a certain $p_{tx}$ value, referred to as *synchronization point (SP)*, almost all transmissions start and finish simultaneously (because of the assumed constant frame duration $L$). As simultaneous transmissions lead to severe co-channel interference and very low system goodput, i.e. $G$ calculated in Section D.3, in this study the $p_{tx}$ value of SP is selected to be that value where the system goodput $G$ reduces below 0.001. By comparing the subplots (a) with (b) in Figure 13, one can find a lower SP with a larger $R$ value, given the same $L$ value.

*A.3 Time Metrics*

Simulation results for the expected value of $T_I$, $T_{RB}$, and $T_{RX}$, shown in Figure 14, again confirm the accuracy of the analytical model. The shapes of $\overline{T_{RB}}$ and $\overline{T_I}$ curves are other witnesses of "synchronized" CSMA transmissions at high $p_{tx}$ values, where the value of $\overline{T_{RB}}$ first increases with increased $p_{tx}$ and then decreases after the maximum is reached, as discovered in section III. Beyond the SP, $\overline{T_{RB}}$ converges to the fixed value of $L$, because almost all transmissions become synchronized. The value of $\overline{T_I}$ with increased $p_{tx}$ reduces and finally converges to 1 in both scenario parameter settings.

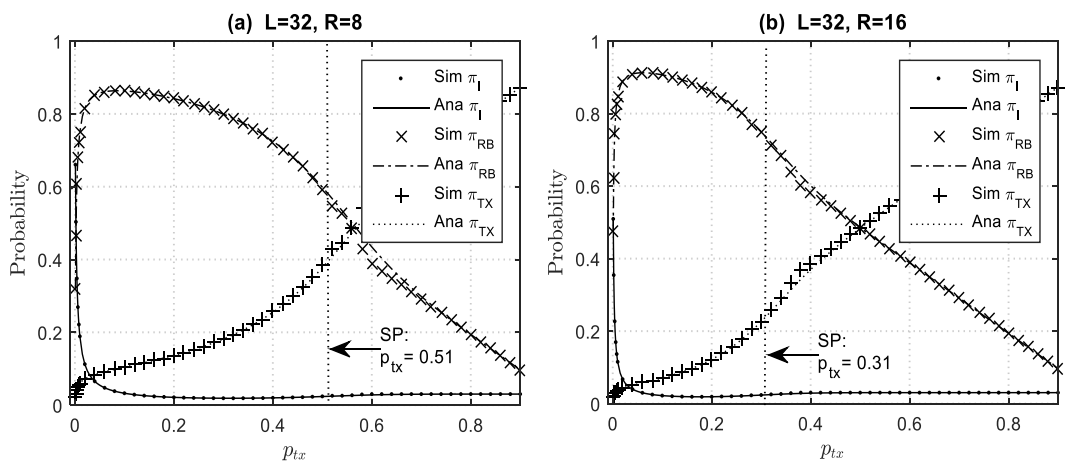

Figure 13: Simulated and analytical limiting probabilities of station's states for (a) L=32, R=8 and (b) L=32, R=16

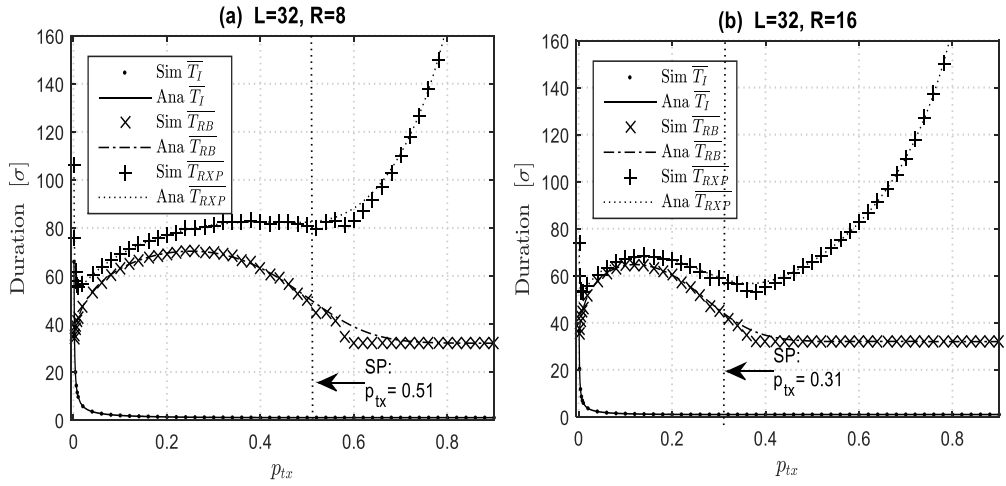

Figure 14: Simulated and analytical results of the mean value of time metrics in scenario (a) L=32, R=8 and (b) L=32, R=16

$\overline{T_{RXP}}$ provides the mean duration between two consecutive frame reception events. From Figure 17, it can be seen that the smallest possible (optimal) value of $\overline{T_{RXP}}$ is reached before $\overline{T_{RB}}$ reaches its maximum. In contrast, too high or too low values of $p_{tx}$ always results in high $\overline{T_{RXP}}$, in both scenarios. The quickly growing value of $\overline{T_{RXP}}$ after passing the SP is due to the high probability of transmission state $TX$ as shown in Figure 13.

The HS problem is more severe in a dense network, i.e. with a higher $R$ value. At a low $p_{tx}$ value, the effect appears as the prolonged $\overline{T_{RB}}$ value, whereas at a high value of $p_{tx}$ the effect appears as the convergence of $\overline{T_{RB}}$ towards $L$, as shown in Figure 14.

*A.4 Interference-Free Probability of Reception Burst and System Goodput G*

The simulated results and the analytical results for $p_{IF}$ and $G$ match very well with both parameter settings, as shown in Figure 15. The value of $p_{IF}$ monotonically decreases from close to 1 to almost 0 with increased $p_{tx}$, whereas the best efficiency of CSMA broadcast can only be achieved with low $p_{tx}$, and the value of $G$ drops dramatically when $p_{tx}$ goes higher. The very low values of $p_{IF}$ and $G$ beyond the SP are because that the synchronized CSMA transmissions are overlapped at almost all receiving stations. The optimal value of $p_{tx}$ for the best performance of $G$ is scenario dependent, as analyzed in the next section. It is worth noting that $G$ stands only for the MAC layer efficiency with respect to the interference. The actual frame success probability in a practical system depends also on other factors, e.g. the decoder performance at the physical layer (PHY) in the presence of co-channel interference, and may result in a higher goodput.

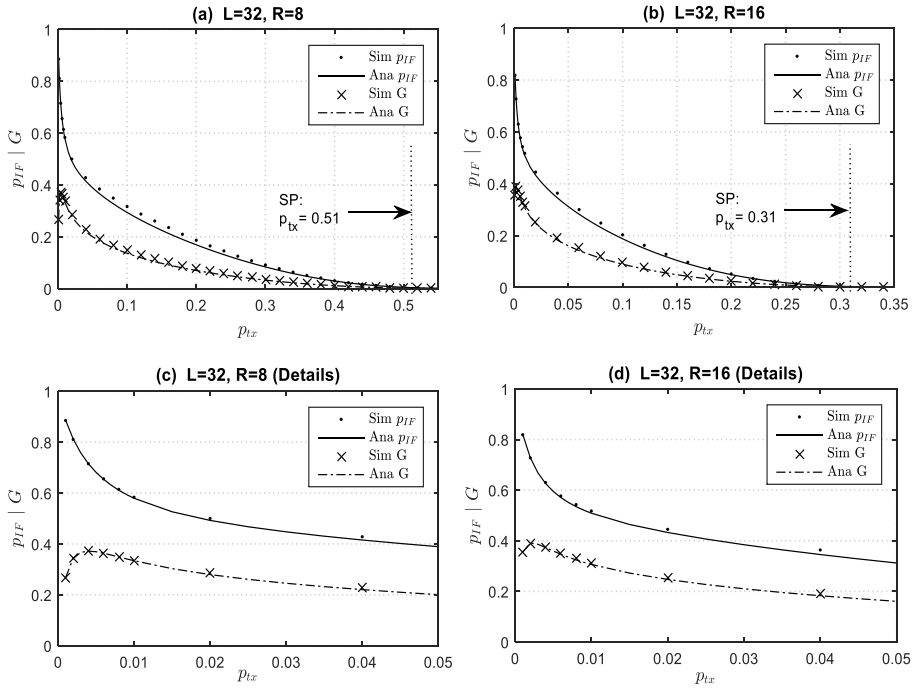

Figure 15: Simulated and analytical results of interference-free probability of reception bursts and system's goodput for (a) L=32, R=8 and (b) L=32, R=16; subplots (c) and (d) show the details at small $p_{tx}$ values in subplots (a) and (b), respectively

In addition to $p_{IF}$, our model also provides the conditional distribution $f_{d_{RX}|IF}$ of the logical distance $d_{RX}$, $1 \leq d_{RX} \leq R$, between a receiver of a frame and its transmitter, conditioned the reception burst is free from any interference. Figure 17 shows the analytical results of $f_{d_{RX}|IF}$ compared to simulated results. As an unbiased rule for all stations in the infinite 1-D network, the interference-free probability of a frame from a closer neighbor is higher than that of a frame from a neighbor with a larger $d_{RX}$ value in a HS scenario. This advantage for closer neighbors is more obvious with high $p_{tx}$ values than with low $p_{tx}$ values, for a given L and R setting.

VIII. FURTHER ANALYTICAL RESULTS AND DISCUSSION

Remarkable impacts of R on the system performance are observed. The analytical results of the mean duration of channel busy period $\overline{T_{RB}}$ and the goodput value G at a station with different values of R are shown in Figure 18. One can see that for the same frame duration value L a higher optimal G value and a lower maximal $\overline{T_{RB}}$ value are achieved with larger R at a lower

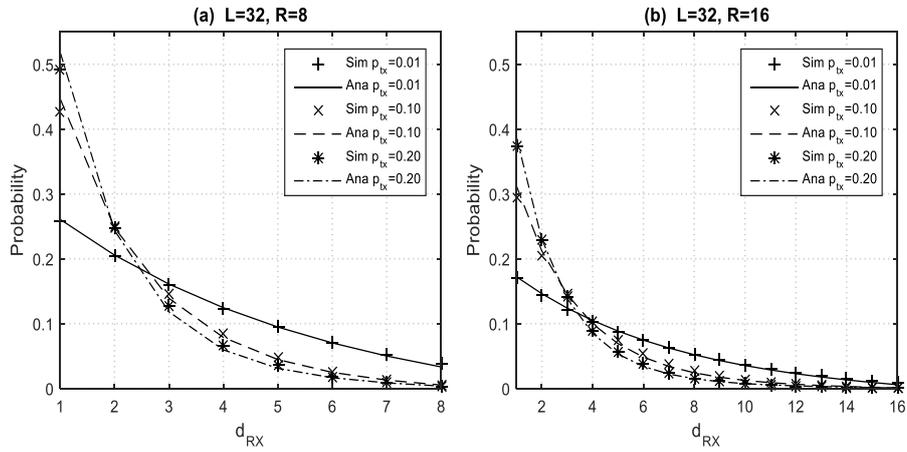

Figure 17: Simulated and analytical conditional distribution of $d_{RX}$ for interference-free reception bursts in scenario (a) L=32, R=8 (b) L=32, R=16

value of $p_{tx}$. However, the larger the value of $R$ is, the faster the value of $G$ drops with increased $p_{tx}$ beyond the maximal value, i.e. the quicker the system becomes "synchronized". It is shown here that the value of $p_{tx}$ is crucial to the system efficiency.

A similar impact on the goodput performance is observed from the frame size $L$, but the effect is less obvious in comparison with the impact of $R$. As shown in Figure 16 (b), a higher optimal value of $G$ is achieved with a higher $L$ value. But increasing the value of $L$ results in a much higher value of $\overline{T_{RB}}$, as shown in Figure 16 (a). Besides, unlike in Figure 18 (b) the values of $G$ for different frame length $L$ reach zero at similar values of $p_{tx}$, as shown in Figure 16 (b).

## IX. CONCLUDING REMARKS

Our modeling and analysis of CSMA with HSs is based on two fundamental assumptions:

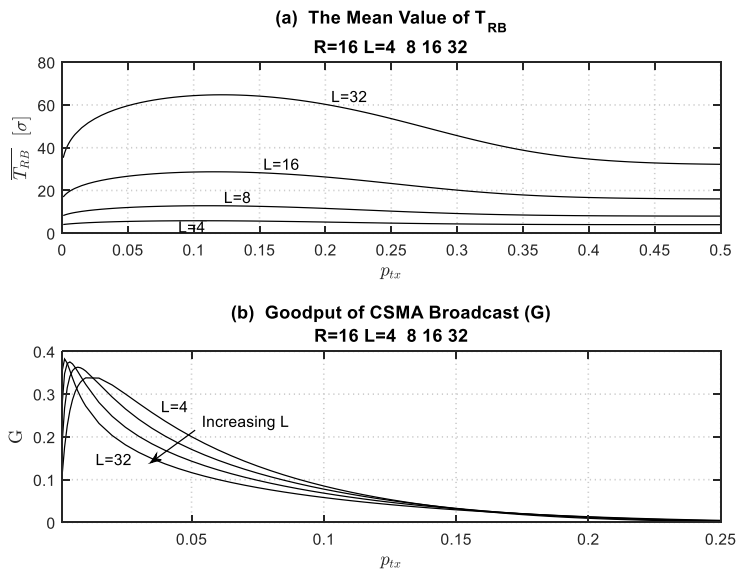

Figure 16: Impact of the frame duration ($L$) on (a) the mean receiving busy time and (b) the system goodput

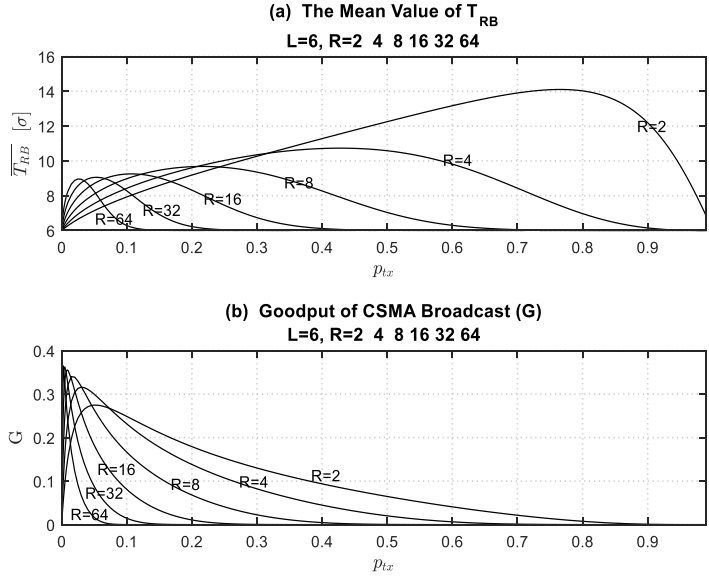

Figure 18: Impacts of the number of one-side neighbors ($R$) on (a) the mean channel busy period and (b) the system goodput

the stationary conditional channel access probability $p_{tx}$ of a station, and the conditional independence among stations in isolated sub-graphs resulting from the clear-channel operation. A HS model is developed for the infinite 1-D CSMA broadcast network. The model takes $p_{tx}$, frame length $L$, and the number of stations $R$ in one-side channel sensing range as input parameters and gives closed-form solutions to steady-state performance metrics, such as the distribution of inter-transmitter distance $d_{TX}$, the mean duration of the channel busy period $T_{RB}$ at a station, the interference-free probability $p_{IF}$ of a reception burst, and the system goodput $G$.

The "double-peak"-shaped PMF curve of $d_{TX}$ and the prolonged $T_{RB}$ due to HSs are perfectly represented in our model and analysis work. Our quantitative results explain why and how with increased $p_{tx}$ the system goodput $G$ first increases and then drops to zero in a CSMA HS network. We also discover that the maximum system goodput $G$ for CSMA broadcast networks increases with the number of stations in the channel sensing range, represented by parameter $R$. But the system becomes more sensitive to the value of $p_{tx}$ for reaching the optimum system goodput with a higher $R$ value. Furthermore, for a given HS scenario, this model also describes how $p_{IF}$ for an interference-free reception burst deteriorates with increased topological distance $d_{RX}$ between transmitter and receiver, as an unbiased rule for every station in the infinite 1-D network.

One of the open problems is how to model the HS problem in generic 2-D networks using the methodology developed in this work. It is shown based on a simulation study in Appendix A of [23] that the assumption of conditional independence still holds in randomly generated 2-D CSMA networks, but a closed-form solution for performance parameters for the HS problem in arbitrary 2-D networks is still missing.

In [25] our HS model for infinite 1-D network is applied to the IEEE 802.11p protocol for evaluating the performance of the co-operative awareness service in a realistic highway scenario with given parameters $L$, $R$, frame generation frequency $\lambda_F$, and the minimum contention window size $CW_{min}$ of IEEE 802.11p.

# ANNEX A

This annex includes Section 5.3.3 from reference [23] on mathematical derivations of the distribution of inter-transmitter distance $d_{TX}$ in the manuscript. This annex corresponds to footnote 3 in the manuscript.

### 5.3.3 Distribution of Inter-Transmitter Distance

The probability mass function $f_{d_{TX}}(k)$, $k = 1,2,3,...$, of inter-transmitter distance $d_{TX}$ in infinite 1-D scenario is calculated for three ranges: Range I: $1 \leq d_{TX} \leq R$, Range II: $R + 1 \leq d_{TX} \leq 2R + 1$, and Range III: $d_{TX} \geq 2R + 2$.

#### 5.3.3.1 Range I: $1 \leq d_{TX} \leq R$

$1 \leq d_{TX} \leq R$ happens only when two or more stations in the same R-Zone transmit simultaneously, whose probability is determined by the number of stations in this R-Zone that sense the channel idle, i.e. the value of $d_F$ at the last idle time slot before these stations start transmitting. Given the geometric distribution of $d_F$ and the stationary conditional channel access probability assumption we have:

$$f_{d_{TX}}(k) = \Pr\{d_{TX} = k\}$$
$$= (1 - p_{tx})^{k-1} \cdot p_{tx} \cdot (1 - p_{O,F})^k, 1 \leq k \leq R \quad (5.5)$$

#### 5.3.3.2 Range II: $R+1 \leq d_{TX} \leq 2R+1$

The distribution of inter-transmitter distance $R + 1 \leq d_{TX} \leq 2R + 1$ is determined by the situations in the R-Zone located right next to the location $n = 2R$, as depicted in Figure 5.8.

Figure 5.9 shows a 2-D Markov chain modeling the situation of the R-Zone shown in Figure 5.8, where the state $NT$ stands for the situation that no station in the R-Zone transmits, i.e. all stations in the R-Zone sense the channel as idle. States $T(i,j)$, $i = 0,1,...R$, $j = 1,2,...,L$ represents the situation that at least one station in the R-Zone is transmitting and the closest transmitter is at the spatial location $i$ in the R-Zone, and the transmitter is transmitting the $j$-th slot of the total frame length $L$ at the current time slot. Spatial locations 0 and $R$ in the R-Zone are indicated in Figure 5.8. Transitions in this Markov chain take place at the starting border of each time slot.

All non-zero transition probabilities of the R-Zone Markov chain are depicted in Figure 5.9 and calculated as follows:

$p_{NT,NT}$ is the probability that no station in the R-Zone starts to transmit in the next slot, conditioned on that no station in the R-Zone is transmitting in the current slot. The number of stations in the R-Zone that sense channel as idle at a given time slot is determined by the value of $d_F$. Given the geometric distribution of $d_F$, $p_{NT,NT}$ is calculated as:

$$p_{NT,NT} = \frac{\sum_{n=1}^{\infty} \Pr\{no\ tx\ in\ \min(R+1, d_F) \mid d_F = n\} \cdot \Pr\{d_F = n\}}{\Pr\{d_F \geq 1\}}$$

$$= \frac{\sum_{n=1}^{R+1}(1-p_{tx})^n \cdot (1-p_{O,F})^n \cdot p_{O,F} + \sum_{n=R+2}^{\infty}(1-p_{tx})^{R+1} \cdot (1-p_{O,F})^n \cdot p_{O,F}}{1 - p_{O,F}}$$

$$= 1 - p_{tx} \cdot \frac{1 - a^{(R+1)}}{1 - a} \qquad (5.6)$$

Where,

$$a = (1 - p_{tx}) \cdot (1 - p_{O,F}) \qquad (5.7)$$

The probability $\Pr\{d_F \geq 1\}$ in the denominator of Eq.(5.6) is introduced due to the condition that no station is transmitting in the R-Zone at the current slot.

$p_{T(i,1),NT}$, $i = 0, 1, \ldots, R$ is the probability that the closest transmitter in the R-Zone in the next time slot is at location $i$, conditioned on no station transmits in the R-Zone at the current time slot. This probability is calculated as:

$$p_{T(i,1),NT} = \frac{\Pr\{the\ closest\ transmitter\ is\ at\ i\ in\ R - Zone \mid d_F \geq i+1\} \cdot \Pr\{d_F \geq i+1\}}{\Pr\{d_F \geq 1\}}$$

$$= \frac{(1-p_{tx})^i \cdot p_{tx} \cdot (1-p_{O,F})^{i+1}}{1 - p_{O,F}} = p_{tx} \cdot a^i \qquad (5.8)$$

Where, $a$ is given in Eq.(5.7).

It can be proved that

$$p_{NT,NT} + \sum_{i=0}^{R} p_{T(i,1),NT} = 1 \qquad (5.9)$$

All other non-zero transition probabilities are 1 in the R-Zone Markov chain, because all transmissions in a R-Zone are synchronized according to the definition of R-Zone in Section 4.4.

$$p_{T(i,j+1),T(i,j)} = 1; \quad i = 0, 1, \ldots, R; \quad j = 1, 2, \ldots, L-1 \qquad (5.10)$$

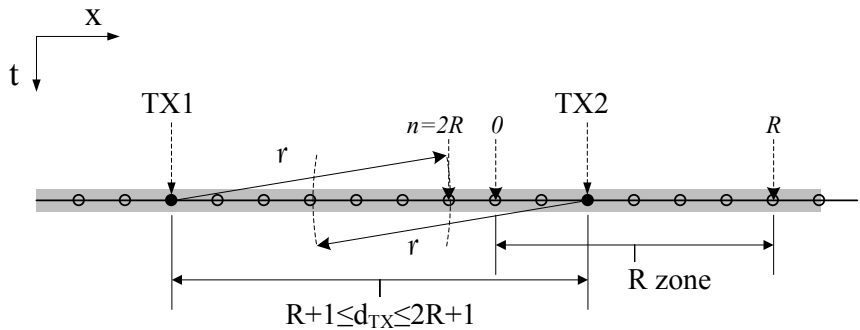

Figure 5.8: Situation of inter-transmitter distance $d_{TX}$ between $R+1$ and $2R+1$

$$p_{NT,T(i,L)} = 1; \quad i = 0,1,\ldots,R \tag{5.11}$$

We denote the limiting distribution of states $NT$ and $T(i,j)$ as $\pi_{NT}$ and $\pi_{T(i,j)}$, $i = 0,1,\ldots,R$; $j = 1,2,\ldots,L$, respectively, and express the limiting distribution of each state using $\pi_{NT}$ and the transition probabilities:

$$\begin{cases} \pi_{T(i,1)} = \pi_{NT} \cdot p_{T(i,1),NT} & , i = 0,1,\ldots,R \\ \pi_{T(i,j)} = \pi_{T(i,j-1)} \cdot p_{T(i,j),T(i,j-1)} & , i = 0,1,\ldots,R; j = 2,3,\ldots,L \end{cases} \tag{5.12}$$

By imposing the normalization condition:

$$\pi_{NT} + \sum_{i=0}^{R}\sum_{j=1}^{L} \pi_{T(i,j)} = 1 \tag{5.13}$$

We have the limiting distribution of the state $NT$ expressed in $p_{0,F}$, $p_{tx}$, $L$, and $R$:

$$\pi_{NT} = \frac{1}{1 + L \cdot p_{tx} \cdot \frac{1-a^{R+1}}{1-a}} \tag{5.14}$$

Where, $a$ is given in Eq.(5.7).

The limiting probabilities of states $T(i,j)$ are derived from Eq.(5.8), Eq.(5.10), Eq.(5.11), Eq.(5.12), and Eq.(5.14):

$$\pi_{T(i,j)} = \frac{p_{tx} \cdot a^i}{1 + L \cdot p_{tx} \cdot \frac{1-a^{R+1}}{1-a}} , i = 0,1,\ldots,R, \ j = 1,2,\ldots,L \tag{5.15}$$

$\pi_{NT}$ is exactly the transition probability $p_{F,O(2R)}$ in the space-domain Markov chain, i.e. the

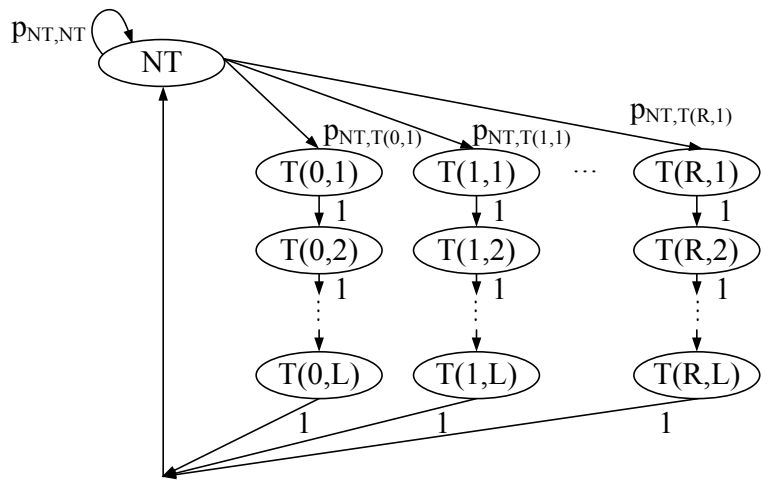

Figure 5.9: 2-D Markov chain for inter-transmitter distance $d_{TX}$ between $R+1$ and $2R+1$

probability that no transmission in the R-Zone, see Figure 5.7.

Besides, the transition probabilities $p_{O(n),O(2R)}$, $n = 0,1,...,R$ in the space-domain Markov chain are exactly the probabilities that the R-Zone is in any of the states $\pi_{T(R-n,j)}$, $j = 1,2,...,L$, i.e.

$$p_{O(n),O(2R)} = \sum_{j=1}^{L} \pi_{T(R-n,j)} = L \cdot \pi_{T(R-n,j)}$$

$$= \frac{L \cdot p_{tx} \cdot a^{R-n}}{1 + L \cdot p_{tx} \cdot \frac{1-a^{R+1}}{1-a}}, n = 0,1,...,R \quad (5.16)$$

Where, $a$ is given in Eq.(5.7). The derivation of (5.16) uses the deduced assumption from the conditional independence assumption that the starting times of transmissions from stations A and B are independent, if A and B do not belong to a same R-Zone and no other station is transmitting between A and B.

Therefore, the distribution of $d_{TX}$ in range $R + 1 \leq d_{TX} \leq 2R + 1$ is given as:

$$f_{d_{TX}}(k) = \Pr\{d_{TX} = k\}$$

$$= \Pr\{d_{TX} \geq R + 1\} \cdot \frac{L \cdot p_{tx} \cdot a^{k-(R+1)}}{1 + L \cdot p_{tx} \cdot \frac{1-a^{R+1}}{1-a}}, R + 1 \leq k \leq 2R + 1 \quad (5.17)$$

Where, $a$ is given in Eq.(5.7), and based on Eq.(5.5), we have:

$$\Pr\{d_{TX} \geq R + 1\} = 1 - \sum_{k=1}^{R} \Pr\{d_{TX} = k\} \quad (5.18)$$

### 5.3.3.3 Range III: $d_{TX} \geq 2R+2$

The relation between $d_F$ and $d_{TX}$ is:

$$d_F = d_{TX} - (2R + 1), \quad d_{TX} \geq 2R + 2 \quad (5.19)$$

Given the geometric distribution of $d_F$ in Eq.(5.2), the distribution of $d_{TX}$ conditioned on $d_{TX} \geq 2R + 2$ is also a geometric distribution of the same parameter $p_{O,F}$:

$$f_{d_{TX}}(k) = \Pr\{d_{TX} = k\}$$

$$= \Pr\{d_{TX} \geq 2R + 2\}(1 - p_{O,F})^{k-2R-2} \cdot p_{O,F}, \quad k \geq 2R + 2 \quad (5.20)$$

Where, according to Eq.(5.14) and Eq.(5.18),

$$\Pr\{d_{TX} \geq 2R + 2\} = \Pr\{d_{TX} \geq R + 1\} \cdot \pi_{NT}$$

$$= \left(1 - \sum_{k=1}^{R} \Pr\{d_{TX} = k\}\right) \cdot \frac{1}{1 + L \cdot p_{tx} \cdot \frac{1 - a^{R+1}}{1 - a}} \qquad (5.21)$$

Where, $\Pr\{d_{TX} = k\}, 1 \leq k \leq R$ and $a$ are from Eq.(5.5) and Eq.(5.7), respectively.

ANNEX B

Section 5.5.3 from reference [23] on mathematical derivations of time metrics is detailed in this annex. For the completeness of the mathematical calculation, we provide the figure and some equations that are already included in the manuscript. This annex corresponds to footnote 4 in the manuscript.

### 5.5.3 Time Metrics

Figure 5.11[1] illustrates time metrics of the CSMA broadcast protocol measured at a station in the infinite 1-D scenario. These metrics are calculated using the time-domain Markov chain.

$T_I$ is the length of the *channel idle period*. As shown in Figure 5.4[2], state $I$ has the fixed self-transition probability $p'_{I|I}$. Thus, the duration that a station stays in state $I$ follows a geometric distribution with parameter $1 - p'_{I|I}$ in this model. Therefore, the expected length of $T_I$ is

$$\overline{T_I} = \sum_{n=1}^{\infty} n \cdot (p'_{I|I})^n \cdot (1 - p'_{I|I}) = \frac{1}{1 - p'_{I|I}} \qquad (5.25)$$

$T_{NI}$ is the length of the *non-idle period* between two consecutive channel idle period. This non-idle period may be either a transmission period $T_{TX}$, when the station is transmitting, or a receiving busy period $T_{RB}$, when the station is receiving. In order to solve the expected value of $T_{NI}$, we assume the occurrence of event "channel goes from non-idle to idle" is a renewal process, i.e. the duration between the starting time of two consecutive idle periods, i.e. $T_I + T_{NI}$, is an i.i.d. random variable. Given the expected length of channel idle period $\overline{T_I}$ in Eq.(5.25), we can calculate the expected value of $T_{NI}$:

$$\overline{T_{NI}} = \overline{T_I} \cdot \frac{1 - \pi_I}{\pi_I} \qquad (5.26)$$

Where, $\pi_I$ is the limiting probability of state $I$ in the time domain Markov chain.

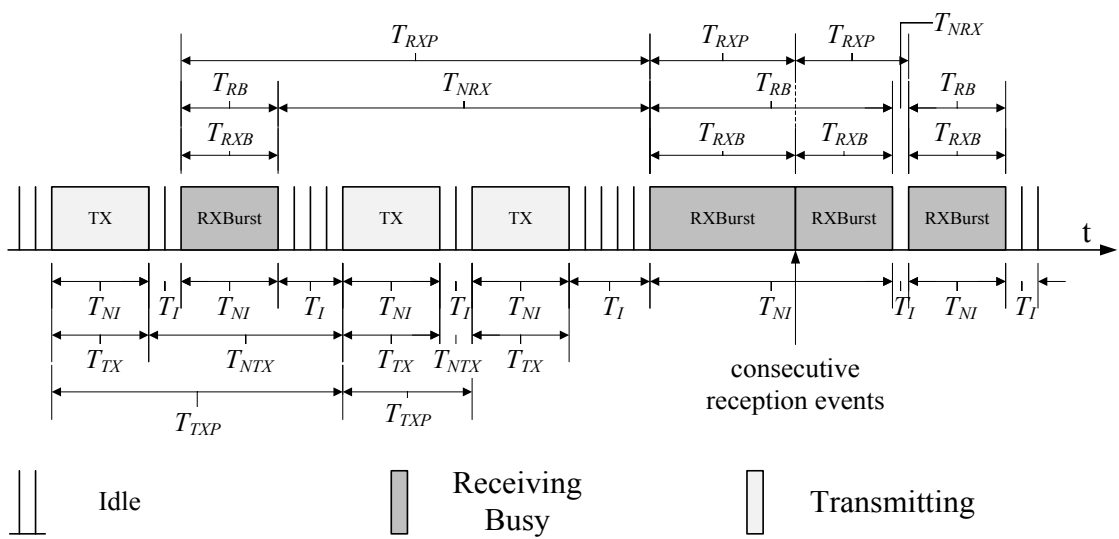

Figure 5.11: Time metrics in hidden station model

---

[1] Figure 5.11 in [23] is numbered as Figure 10 in the manuscript.
[2] Figure 5.4 in [23] is numbered as Figure 6 in the manuscript.

$T_{TX}$ is the duration of a frame transmission. $T_{TX}$ is always $L$.

$T_{NTX}$ is the duration measured from the ending time of a frame transmission to the starting time of the next frame transmission at a station. Assuming the occurrence of event "the station starts a frame transmission" is a renewal process, the expected value of $T_{NTX}$ is calculated as:

$$\overline{T_{NTX}} = T_{TX} \cdot \left( \frac{1}{\sum_{n=1}^{L} \pi_{TX_{(L,n)}}} - 1 \right) \tag{5.27}$$

Where, $\pi_{TX_{(L,n)}}, 1 \leq n \leq L$, is the limiting probability of state $TX_{(L,n)}$ in the time domain Markov chain.

$T_{TXP}$ is the length of a *transmission period*, which is defined as the time between the starting point of two consecutive frame transmissions. The expected value of $T_{TXP}$ is:

$$\overline{T_{TXP}} = T_{TX} + \overline{T_{NTX}} \tag{5.28}$$

Where, $\overline{T_{NTX}}$ is solved in Eq.(5.27).

$T_{RB}$ is the length of a *receiving busy period*[3], during which a station constantly senses the channel busy. It may consist of one or more consecutive reception bursts, as shown in Figure 5.11. In this scenario, $T_{NI}$ is either a transmission period $T_{TX}$ or a receiving busy period $T_{RB}$ following a channel idle period $T_I$. Therefore, the expected length of $T_{NI}$ is also calculated as

$$\overline{T_{NI}} = \frac{1 - p'_{I|I} - p'_{TX|I}}{1 - p'_{I|I}} \cdot \overline{T_{RB}} + \frac{p'_{TX|I}}{1 - p'_{I|I}} \cdot \overline{T_{TX}} \tag{5.29}$$

Where, $\frac{1 - p'_{I|I} - p'_{TX|I}}{1 - p'_{I|I}}$ is the probability that a $T_I$ period is followed by a $T_{RB}$ period, whereas $\frac{p'_{TX|I}}{1 - p'_{I|I}}$ is the probability that a $T_I$ period is followed by a $T_{TX}$ period. From Eq.(5.29), the expected length of a receiving busy period $T_{RB}$ is given as:

$$\overline{T_{RB}} = \frac{(1 - p'_{I|I}) \cdot \overline{T_{NI}} - p'_{TX|I} \cdot L}{1 - p'_{I|I} - p'_{TX|I}} \tag{5.30}$$

Where, $\overline{T_{NI}}$ is given in Eq.(5.26).

$T_{RXB}$ is the length of a *reception burst*, which is defined as the duration from the time a station starts a reception, i.e. the time the station enters any of the following states in the time-domain Markov chain: $B_{(L,1)}$, $V_{(L,1)}$, $VBE_{(L,1)}$, and $VBL_{(n,1)}, 1 \leq n \leq L - 1$, to the time when the station detects the end of the signal of the received frame or the end of all signals of the received overlapping frames, i.e. the time that the end of any of the following states in the time-domain

---

[3] $T_{RB}$ is referred to as *Channel Busy Period* in the manuscript

Markov chain is reached at the receiving station: $B_{(L,L)}$, $V_{(L,L)}$, and $VBE_{(L,L)}$. The length of a reception burst may be longer than the frame length $L$ due to the overlapped transmission from hidden stations. Reception bursts are marked with "RXBurst" in Figure 5.11. The introduction of period $T_{RXB}$ is due to the consecutive reception events, which may happen when the station is in state $V_{(L,L)}$. There is a probability $p'_{B|V}$ that the station goes on with a new reception without going back to the idle state, as shown in Figure 5.11. The next frame is received from an unsynchronized transmitter that is hidden to the previous transmitter. Given $\pi_{B_{(L,L)}}$, $\pi_{V_{(L,L)}}$, and $\pi_{VBE_{(L,L)}}$ are the limiting probability of states $B_{(L,L)}$, $V_{(L,L)}$, and $VBE_{(L,L)}$, respectively, we calculate the probability that a consecutive reception event happens, conditioned on the current reception burst is over:

$$p_{ConRX} = \frac{\pi_{V_{(L,L)}}}{\pi_{B_{(L,L)}} + \pi_{V_{(L,L)}} + \pi_{VBE_{(L,L)}}} \cdot p'_{B|V} \qquad (5.31)$$

Therefore, the probability that the channel returns idle after a reception burst is $1 - p_{ConRX}$. Assuming the probability $p_{ConRX}$ is stationary, it is observed that the number of reception bursts $T_{RXB}$ in one receiving busy period $T_{RB}$ follows the geometric distribution with parameter $1 - p_{ConRX}$. Therefore, we have:

$$\overline{T_{RB}} = \overline{T_{RXB}} \cdot \sum_{n=1}^{\infty} n \cdot (p_{ConRX})^{n-1} \cdot (1 - p_{ConRX}) = \frac{\overline{T_{RXB}}}{1 - p_{ConRX}} \qquad (5.32)$$

And:

$$\overline{T_{RXB}} = \overline{T_{RB}} \cdot (1 - p_{ConRX}) \qquad (5.33)$$

Where, $\overline{T_{RB}}$ is given in Eq.(5.30).

$T_{NRX}$ is the length of *non-reception period*, in which the station may be sensing the channel as idle or transmitting a frame, as shown in Figure 5.11. Assuming the occurrence of receiving busy periods is a renewal process, the expected length of $T_{NRX}$ is calculated as:

$$\overline{T_{NRX}} = \frac{\pi_I + \sum_{n=1}^{L} \pi_{TX_{(L,n)}}}{1 - \left(\pi_I + \sum_{n=1}^{L} \pi_{TX_{(L,n)}}\right)} \cdot \overline{T_{RB}} \qquad (5.34)$$

Where, $\pi_{TX_{(L,n)}}$ is the limiting distribution of states $TX_{(L,n)}, 1 \leq n \leq L$ in the time-domain Markov chain, and $\overline{T_{RB}}$ is given in Eq.(5.30).

$T_{RXP}$ is the length of the *reception period*, which is defined as the duration between the starting time of two consecutive reception bursts. See Figure 5.11. Assuming that the occurrence of frame reception burst is a renewal process, the expected length of reception period $T_{RXP}$ is expressed as:

$$\overline{T_{RXP}} = (1 - p_{ConRX}) \cdot (\overline{T_{RXB}} + \overline{T_{NRX}}) + p_{ConRX} \cdot \overline{T_{RXB}} \qquad (5.35)$$

Where, $\overline{T_{RXB}}$ is given in Eq.(5.33), $\overline{T_{NRX}}$ is given in Eq.(5.34), and $p_{ConRX}$ is from Eq.(5.31).

Among these time metrics $\overline{T_I}, \overline{T_{TXP}}, \overline{T_{RB}},$ and $\overline{T_{RXP}}$ are of particular interest to this study.

ANNEX C

Section 5.5.4 from reference [23] on mathematical derivations of the interference-free probability $p_{IF}$ of reception burst is provided in this annex. This annex corresponds to endnote 5 in the manuscript.

### 5.5.4 Interference-Free Probability of Reception Burst

As a metric of the reliability of CSMA broadcast in hidden station scenarios, we define the interference-free probability $p_{IF}$ of a reception burst as the probability that a frame is received without being interfered by any other transmission, conditioned on the receiver starts a reception burst. It is worth mentioning that according to the definition of reception burst, an interference-free reception burst consists of only one frame, whereas a reception burst consisting of two or more frames is always interfered. $p_{IF}$ is also the ratio of reception bursts free from interference out of all reception bursts at a receiver. Particularly, we are interested in the $p_{IF}$ performance with respect to a given topological reception distance $d_{RX}$ from the receiver to the transmitter.

To calculate the interference probability $p_{IF}$ two kinds of interference have to be considered: a) the same-side interference, which is generated by transmissions from stations at the same side as the intended transmitter, b) the other-side interference, which is generated by transmissions from stations at the other side of the intended transmitter. A received frame is interference-free only when it is neither affected by the same-side interference nor affected by the other-side interference in any of the $L$ time slots during the reception, where $L$ is the frame length. As a deduction from the conditional independence assumption, the occurrence of same-side interference is independent from that of other-side interference. It is worth mentioning that owing to the assumptions of fixed frame length and identical channel sensing range at all stations, transmissions generating same-side interference are always synchronized to the intended transmission, whereas the transmissions generating other-side interference may be unsynchronized to the intended transmission, e.g. in case of the hidden station problem.

Figure 5.12 illustrates the reception distance $d_{RX}$ from a receiver to the intended transmitter of a frame, e.g. $TX1$. In Figure 5.12, the receiver is at the right side of $TX1$ and $d_{RX} = 3$. It also shows the situation that a transmitter $TX1'$ at the left side of $TX1$ generates same-side interference to receivers having $d_{RX} = 1$ and $d_{RX} = 2$ at the right side of $TX1$. In the lower part of Figure 5.12, receivers having $d_{RX} = 4$, $d_{RX} = 5$, and $d_{RX} = 6$ at the right side of $TX1$, which fall into the area $B$ between $TX1$ and $TX2$, suffer from the other-side interference due to the later started transmission from $TX2$.

The probability $p_{IFS}$ that a receiver is free from same-side interference depends only on the reception distance $d_{RX}$ and the distribution of $d_{TX}$. For a given $d_{RX}$, $p_{IFS}$ is calculated as:

$$p_{IFS}(d_{RX}) = \Pr\{d_{TX} \geq R - d_{RX} + 1\} \ , 1 \leq d_{RX} \leq R \qquad (5.36)$$

Where, $\Pr\{d_{TX} \geq R - d_{RX} + 1\}$ is calculated using the distribution of $d_{TX}$ given in Section 5.3.3.

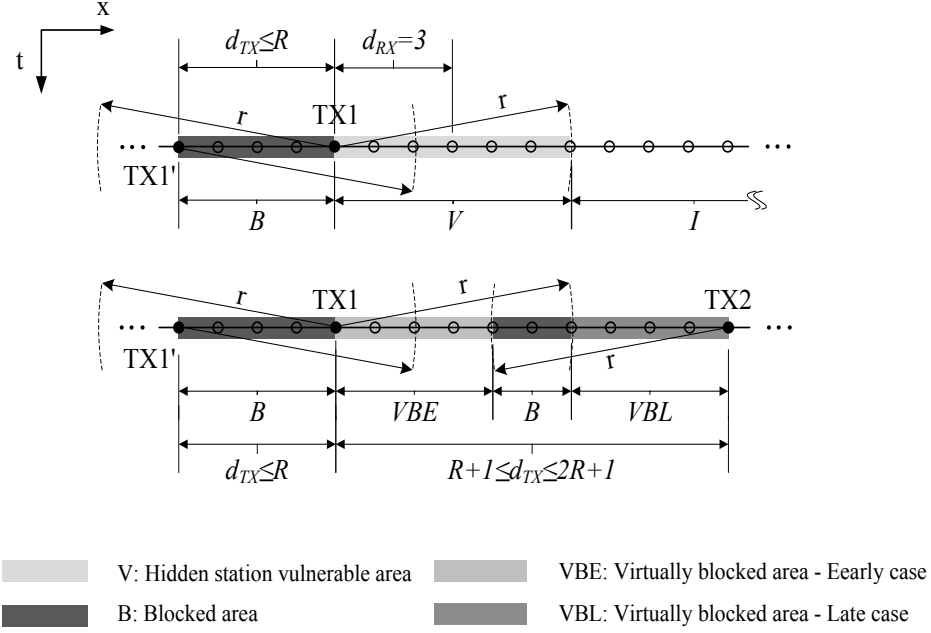

V: Hidden station vulnerable area
B: Blocked area
VBE: Virtually blocked area - Eearly case
VBL: Virtually blocked area - Late case

Figure 5.12: Frame reception distance $d_{RX}$ at the right side of transmitter *TX1*

In order to receive a frame free from other-side interference, a receiver can only stay in areas $V$, $VBE$, or $VBL$ throughout the reception of a frame. Area $V$ always contains $R$ stations, whereas $VBE$ and $VBL$ areas may contain 1 to $R$ stations depending on the inter-transmitter distance $d_{TX}$, where $R + 2 \leq d_{TX} \leq 2R + 1$.

Conditioned on a receiver is in either a $VBE$ or a $VBL$ area, the distribution of $d_{RX}$ from this receiver to the closest transmitter is calculated using the inter-transmitter distance $d_{TX}$ in Section 5.3.3 and the Bayes' Theorem [24]:

$$f_{d_{RX}|VB}(k) = \Pr\{d_{RX} = k | receiver\ is\ in\ either\ VBE\ or\ VBL\ area\}$$

$$= \frac{\Pr\{VBE\ or\ VBL\ area\ contains\ k\ or\ more\ stations\}}{\sum_{j=1}^{R} \Pr\{VBE\ or\ VBL\ area\ contains\ j\ or\ more\ stations\}}$$

$$= \frac{\Pr\{R + k + 1 \leq d_{TX} \leq 2R + 1\}}{\sum_{j=1}^{R} \Pr\{R + j + 1 \leq d_{TX} \leq 2R + 1\}}, 1 \leq k \leq R \quad (5.37)$$

If the receiver is in area $V$, $d_{RX}$ is uniformly distributed in the range from 1 to $R$:

$$f_{d_{RX}|V}(k) = \frac{1}{R}, 1 \leq k \leq R \quad (5.38)$$

To calculate the probability $p_{IF}$ we refer to the time-domain Markov chain of the receiver. As shown in Figure 5.4[4], $I$ and $V_{(L,L)}$ are the only states that a station can start a frame reception. The probability that a station starts a frame reception is:

$$p_{RX} = \pi_I \cdot \left(1 - p'_{I|I} - p'_{TX|I}\right) + \pi_{V_{(L,L)}} \cdot p'_{B|V} \quad (5.39)$$

---
[4] Figure 5.4 in [23] is numbered as Figure 6 in the manuscript.

Conditioned on the event that a station starts a frame reception, the probability that the frame is received interference-free equals the probability that state $V_{(L,L)}$ or state $VBE_{(L,L)}$ is reached in $L$ steps without entering any of states $B_{(l,n)}$, where $1 \leq l \leq L$ and $1 \leq n \leq l$, and the frame is free from the same-side interference.

Here, three cases are identified for calculating $p_{IF}$:

**Case I**: The frame reception starts with state $V_{(L,1)}$. The probability of this event is:

$$p_{IF\_I} = \pi_I \cdot \left( p'_{V|I} + \frac{p'_{VBL|I}}{L} \right) + \pi_{V(L,L)} \cdot p'_{B|V} \tag{5.40}$$

In this case, the probability that state $V_{(L,L)}$ is reached in $L$ steps at the receiver with distance $d_{RX}$ to the frame transmitter is:

$$p_{IF\_I\_V}(d_{RX}) = \left[ p_{V|V}(d_{RX}) \right]^{L-1}, 1 \leq d_{RX} \leq R \tag{5.41}$$

Where, $p_{V|V}(d_{RX})$ is the transition probability from state $V(L,n)$ to state $V(L, n+1)$, $1 \leq n \leq L-1$, at the receiver with $d_{RX}$. $p_{V|V}(d_{RX})$ is calculated by Eq.(B.52) in Appendix B.9.

The probability that state $VBE_{(L,L)}$ is reached in $L$ steps at the receiver with $d_{RX}$ is:

$$p_{IF\_I\_VBE}(d_{RX}) = p_{VBE|V}(d_{RX}) \cdot \sum_{i=0}^{L-2} \left[ p_{V|V}(d_{RX}) \right]^i, 1 \leq d_{RX} \leq R \tag{5.42}$$

Where, $p_{VBE|V}(d_{RX})$ is the transition probability from state $V(L,n)$ to state $VBE(L, n+1)$, $1 \leq n \leq L-1$, at the receiver with $d_{RX}$. The calculation of this probability is given by Eq.(B.46) in Appendix B.8.

**Case II**: The frame reception starts with state $VBE_{(L,1)}$ at a receiver. The probability of this event is:

$$p_{IF\_II} = \pi_I \cdot p'_{VBE|I} \tag{5.43}$$

In this case, all stations in the $VBE$ area reach state $VBE_{(L,L)}$ with probability 1, as shown in Figure 5.4[5]:

$$p_{IF\_II\_VBE}(d_{RX}) = 1, 1 \leq d_{RX} \leq R \tag{5.44}$$

**Case III**: The frame reception starts with one of states $VBL_{(l,1)}$, $1 \leq l \leq L-1$, with the probability:

---

[5]Figure 5.4 in [23] is numbered as Figure 6 in the manuscript.

$$p_{IF\_III} = \pi_I \cdot \left(\frac{p'_{VBL|I}}{L}\right) \qquad (5.45)$$

In Case III, similar to in Case I, an interference-free reception may end with state $V_{(L,L)}$ or $VBE_{(L,L)}$ in $L$ steps. If the frame reception starts with state $VBL_{(L-1,1)}$, the probability that the state $V_{(L,L)}$ is reached after $L$ steps is 1. Otherwise, if the frame reception starts with state $VBL_{(l,1)}$, $1 \leq l \leq L-2$, the probability that state $V_{(L,L)}$ is reached after $L$ steps at the receiver with $d_{RX}$ is:

$$p_{IF\_III\_V}(l, d_{RX}) = \left[p_{V|V}(d_{RX})\right]^{L-1-l}, 1 \leq l \leq L-1, 1 \leq d_{RX} \leq R \qquad (5.46)$$

Where, $p_{V|V}(d_{RX})$ is given by Eq.(B.52) in Appendix B.9.

The probability that state $VBE_{(L,L)}$ is reached in $L$ steps at the receiver with $d_{RX}$ is:

$$p_{IF\_III\_VBE}(l, d_{RX}) = \begin{cases} p_{VBE|V}(d_{RX}) \cdot \sum_{i=0}^{L-2-l}\left[p_{V|V}(d_{RX})\right]^i & , 1 \leq l \leq L-2, 1 \leq d_{RX} \leq R \\ 0 & , l = L-1, \quad 1 \leq d_{RX} \leq R \end{cases}$$

$$(5.47)$$

Where, $p_{VBE|V}(d_{RX})$ and $p_{V|V}(d_{RX})$ are given in Appendix B.8 and Appendix B.9, respectively.

Finally, the mean interference-free probability $p_{IF}$ of a reception burst is calculated using Eq.(5.36) to Eq.(5.47):

$$p_{IF} = \frac{1}{p_{RX}} \cdot \sum_{d_{RX}=1}^{R} p_{IFS}(d_{RX})$$

$$\cdot \Big\{ p_{IF\_I} \cdot [p_{IF\_I\_V}(d_{RX}) + p_{IF\_I\_VBE}(d_{RX})] \cdot f_{d_{RX}|V}(d_{RX}) + p_{IF\_II} \cdot f_{d_{RX}|VB}(d_{RX})$$

$$+ p_{IF\_III} \cdot \sum_{l=1}^{L-1}[p_{IF\_III\_V}(l, d_{RX}) + p_{IF\_III\_VBE}(l, d_{RX})] \cdot f_{d_{RX}|VB}(d_{RX}) \Big\}$$

$$(5.48)$$

Using Eq.(5.36) to Eq.(5.48), we can calculate the conditional distribution of $d_{RX}$ given a frame is received interference-free:

$$f_{d_{RX}|IF}(k) = \frac{1}{p_{IF}} \cdot \frac{1}{p_{RX}} \cdot p_{IFS}(k)$$

$$\cdot \left\{ p_{IF\_I} \cdot \left[ p_{IF\_I\_V}(k) + p_{IF\_I\_VBE}(k) \right] \cdot f_{d_{RX}|V}(k) + p_{IF\_II} \cdot f_{d_{RX}|VB}(k) + p_{IF\_III} \right.$$

$$\left. \cdot \sum_{l=1}^{L-1} \left[ p_{IF\_III\_V}(l,k) + p_{IF\_III\_VBE}(l,k) \right] \cdot f_{d_{RX}|VB}(k) \right\}, 1 \leq k \leq R$$

(5.49)

ANNEX D

Appendix B from reference [23] on mathematical derivations of supporting probabilities in the time-domain Markov chain is provided in this annex. This annex corresponds to endnote 2 in the manuscript.

# APPENDIX B

# Calculation of Supporting Probabilities in the Time-Domain Markov Chain

The nine supporting probabilities in the time-domain Markov chain, as introduced in Section 5.2.2.2[6], are calculated in this appendix using known parameters $p_{tx}$, $L$, and $R$, as well as the parameter $p_{0,F}$ for the geometric distribution of the size of interference free area $d_F$, as introduced in Section 5.3.1[7].

Note: Based on the assumed homogenous behavior of stations and the uniformly distributed network topology, all probabilities calculated here are for the steady state of the system and applicable to all stations.

## B.1 $p'_{I|I}$

For any station in the infinite 1-D scenario, $p'_{I|I}(d_F, x)$ is the probability that no station in its one-side channel sensing range transmits in the next time slot, conditioned on this station is in a channel free area in this time slot. Assume the station is at the position $x$ in a channel free area of size $d_F$, $d_F \geq 1$, $1 \leq x \leq d_F$. $p'_{I|I}$ is the expected value of $p'_{I|I}(d_F, x)$ over all possible values of $d_F$ and $x$:

$$p'_{I|I} = \sum_{n=1}^{\infty} \sum_{m=1}^{n} p'_{I|I}(d_F, x) \cdot \Pr\{x = m\} \cdot \Pr\{d_F = n\} \quad \text{(B.1)}$$

### B.1.1 Case I: $1 \leq d_F \leq R+1$

As shown in Figure B.1, for all stations in this case $p'_{I|I}$ is identical and calculated as the probability that none of $d_F$ stations starts transmit in the next time slot.

$$\overline{p'_{I|I}(d_F)} = (1 - p_{tx})^{d_F} \quad , 1 \leq d_F \leq R + 1 \quad \text{(B.2)}$$

---

[6]Section 5.2.2.2 of [23] corresponds to Section VI.A.2 in the manuscript.
[7]Section 5.3.1 of [23] corresponds to Section VI.B.1 in the manuscript.

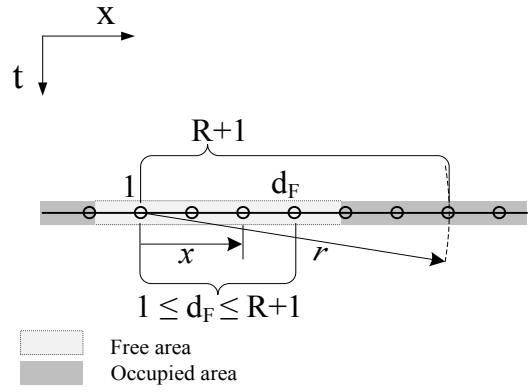

Figure B.1: Example situation of case I

### B.1.2 Case II: $R+2 \leq d_F \leq 2R$

In this case, as shown in Figure B.2, the probability $p'_{I|I}$ for stations at different location, marked as $x$ in the figure, is different. For this reason, three areas, namely $A$, $B$, and $C$, as shown in Figure B.2, are identified for the analysis.

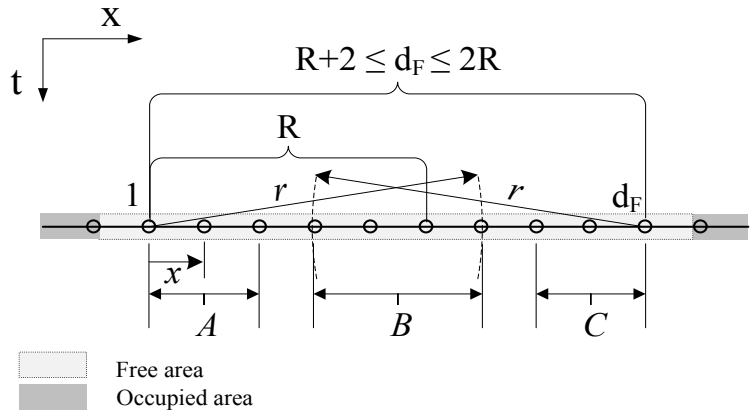

Figure B.2: Example situation of case II

For stations in area $A$, i.e. $1 \leq x \leq d_F - R$, $p'_{I|I}(d_F, x)$ is the probability that no station in the one-side channel sensing range of $x$ in the free area, including the station $x$, starts transmit in the next time slot, i.e.

$$p'_{I|I}(d_F, x) = (1 - p_{tx})^{x+R} \quad , 1 \leq x \leq d_F - R - 1 \tag{B.3}$$

Similarly, for area $B$ and $C$, $p'_{I|I}(x)$ is calculated as

$$p'_{I|I}(d_F, x) = (1 - p_{tx})^{d_F} \quad , d_F - R \leq x \leq R + 1 \tag{B.4}$$

$$p'_{I|I}(d_F, x) = (1 - p_{tx})^{d_F - x + R + 1} \quad , R + 2 \leq x \leq d_F \tag{B.5}$$

The mean value of $p'_{I|I}(d_F, x)$ over all stations in the free area in case II is

$$\overline{p'_{I|I}(d_F)} = \frac{1}{d_F} \cdot \sum_{x=1}^{d_F} p'_{I|I}(d_F, x)$$

$$= \frac{1}{d_F} \cdot \left[ \frac{2}{p_{tx}} \cdot (1-p_{tx})^{R+1} + \left(2 \cdot R - \frac{2 \cdot (1-p_{tx})}{p_{tx}}\right) \cdot (1-p_{tx})^{d_F} - d_F \cdot (1-p_{tx})^{d_F} \right], \quad R+2 \leq d_F \leq 2R$$

(B.6)

### B.1.3 Case III: $d_F \geq 2R+1$

In this case, three areas $A$, $B$, and $C$ are also identified for the analysis, as shown in Figure B.3.

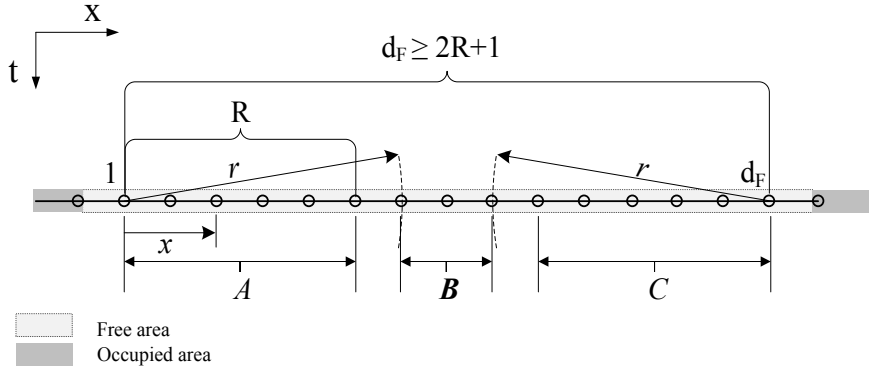

Figure B.3: Example situation of case III

$p'_{I|I}(d_F, x)$ is calculated for each area:

$$p'_{I|I}(d_F, x) = \begin{cases} (1-p_{tx})^{x+R} & 1 \leq x \leq R \\ (1-p_{tx})^{2R+1} & R+1 \leq x \leq d_F - R \\ (1-p_{tx})^{d_F-x+R+1} & d_F - R + 1 \leq x \leq d_F \end{cases} \quad (B.7)$$

The mean value of $p'_{I,I}(d_F, x)$ over all stations in the free area in case III is given as:

$$\overline{p'_{I,I}(d_F)} = \frac{1}{d_F} \sum_{x=1}^{d_F} p'_{I|I}(d_F, x)$$

$$= \frac{1}{d_F} \cdot \left[ \frac{2}{p_{tx}} \cdot (1-p_{tx})^{R+1} - \left(2R + \frac{2}{p_{tx}}\right)(1-p_{tx})^{2R+1} + d_F(1-p_{tx})^{2R+1} \right]$$

$$, d_F \geq 2R+1 \quad (B.8)$$

Given $d_F$ follows the geometric distribution with parameter $p_{0,F}$, $p'_{I|I}$ for the infinite 1-D scenario is calculated according to Eq.(B.1):

$$p'_{I|I} = \frac{\sum_{n=1}^{\infty} \sum_{x=1}^{n} p'_{I|I}(n,x) \cdot \Pr\{d_F = n | d_F \geq 1\}}{\sum_{n=1}^{\infty} n \cdot \Pr\{d_F = n | d_F \geq 1\}}$$

$$= \sum_{n=1}^{\infty} \overline{p'_{I|I}(n)} \cdot n \cdot (1-p_{O,F})^{n-1} \cdot p_{O,F}^2$$

$$= p_{O,F}^2 \cdot (1-p_{tx})$$

$$\cdot \frac{1 - 2 \cdot (R+2) \cdot a^{R+1} + 2 \cdot (R+1) \cdot a^{R+2} + (2R+1) \cdot a^{2R} - 2R \cdot a^{2R+1}}{(1-a^2)} + p_{O,F}^2$$

$$\cdot \left[ 2R - \frac{2 \cdot (1-p_{tx})}{p_{tx}} \right] \cdot (1-p_{tx}) \cdot \frac{a^{R+1} \cdot (1-a^{R-1})}{1-a} + \frac{2 \cdot p_{O,F}}{p_{tx}} \cdot (1-p_{tx})^{R+1}$$

$$\cdot (1-p_{O,F})^{R+1} - \left( 2R + \frac{2}{p_{tx}} \right) \cdot (1-p_{tx})^{2R+1} \cdot p_{O,F} \cdot (1-p_{O,F})^{2R}$$

$$+ (1-p_{tx})^{2R+1} \cdot \left[ (2R+1) \cdot (1-p_{O,F})^{2R} - 2R \cdot (1-p_{O,F})^{2R+1} \right] \quad (B.9)$$

Where,

$$a = (1-p_{tx}) \cdot (1-p_{O,F}) \quad (B.10)$$

## B.2 $p'_{TX|I}$

According to the assumption of stationary conditional channel access probability in Section 3.1.1[8], the probability a station starts to transmit in the next time slot, conditioned on it senses the channel idle in this time slot is:

$$p'_{TX|I} = p_{tx} \quad (B.11)$$

## B.3 $p'_{V|I}$

According to the definition of hidden-station vulnerable area $V$ in Section 5.2.1.1[9], the transition of a station from state $I$ to state $V$ is always caused by the event that at least one neighbor in the channel sensing range of this station starts to transmit in the next time slot. If the situation in the next time slot satisfies the definition of area, i.e. an channel free area with size greater than zero is next to the occupied area, in which the station is located, the station together with other $R-1$ neighbors of it transit to from state $I$ to state $V$.

---

[8]This is Assumption 1 in Section V of the manuscript.
[9]Vulnerable area V is defined in Section VI.A.1 in the manuscript.

Due to the ergodicity property of the infinite 1-D scenario, for a free area with size $d_F$ the mean probability $\overline{p_{V|I}(d_F)}$ that a station in this free area transits from state $I$ to state $V$ can also be calculated as the mean percentage of stations transits from $I$ to $V$ out of $d_F$ stations. To find out this percentage we investigate the situation that the station at location $x$, $(1 \leq x \leq d_F)$ in the channel free area, as shown in Figure B.1, starts to transmit in the next time slot. For a given $d_F$, we have

$$\overline{p_{V|I}(d_F)} = \frac{1}{d_F} \cdot \sum_{x=1}^{d_F} R \cdot p_V(d_F, x) \cdot p_{tx}(x) \tag{B.12}$$

$R$ appears in Eq.(B.12) because a hidden station vulnerable area $V$ always consists of $R$ contiguous stations, as shown in Figure 5.1[10]. $p_V(d_F, x)$ is the probability that the transmission at location $x$ will result in a $V$ area, conditioned on the station at location $x$ starts transmission. $p_{tx}(x)$ is the probability that station at location $x$ starts transmission, which equals to $p_{tx}$ for all stations in this scenario. $p_V(d_F, x)$ is calculated for each location $x$ and $d_F$, as shown in the following subsections.

### B.3.1 Case I: $1 \leq d_F \leq R+1$

It is not possible for a free area with size $1 \leq d_F \leq R + 1$ to generate a $V$ area. Therefore,

$$\overline{p_{V|I}(d_F)} = 0, 1 \leq d_F \leq R + 1 \tag{B.13}$$

### B.3.2 Case II: $R+2 \leq d_F \leq 2R+2$

For the simplicity of analysis, we only investigate the situation at the right side of $x$. Due to the symmetric network topology, the derived $\overline{p_{V|I}(d_F)}$ shall be multiplied by 2 for obtaining the full probability.

As shown in Figure B.4, for $x$ in area $A$, i.e. $1 \leq x \leq d_F - R - 1$, at the right side of $x$, $p_V(d_F, x)$ is the probability that no station starts transmit in the area from $x + 1$ to $d_F$ in the next time slot. Therefore,

$$p_V(d_F, x) = (1 - p_{tx})^{d_F - x}, 1 \leq x \leq d_F - R - 1 \tag{B.14}$$

---

[10] Figure 5.1 in [23] is numbered as Figure 5 in the manuscript.

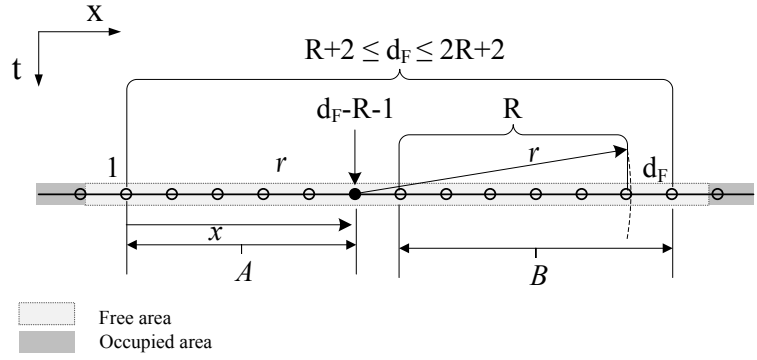

Figure B.4:   Example situation of calculating $p_V$ in case II

For $x$ in area B ($d_F - R \leq x \leq d_F$), as shown in Figure B.4, any transmission in this area will not cause a $V$ area at the right side of $x$. Therefore,

$$p_V(d_F, x) = 0 \ , d_F - R \leq x \leq d_F \tag{B.15}$$

From Eq.(B.13), Eq.(B.14), and Eq.(B.11), we have

$$\overline{p'_{V|I}(d_F)} = \frac{1}{d_F} \sum_{x=1}^{d_F} R \cdot 2 \cdot p_V(d_F, x) \cdot p_{tx}(x) = \frac{2R}{d_F} \cdot [(1 - p_{tx})^{R+1} - (1 - p_{tx})^{d_F}]$$

$$, R + 2 \leq d_F \leq 2R + 2 \tag{B.16}$$

### B.3.3   Case III: $d_F \geq 2R+3$

In this case, three areas $A$, $B$, and $C$, are identified for the calculation of $p_V(d_F, x)$, as shown in Figure B.5.

Similar to Case II, we focus at the situation at the right side of $x$.

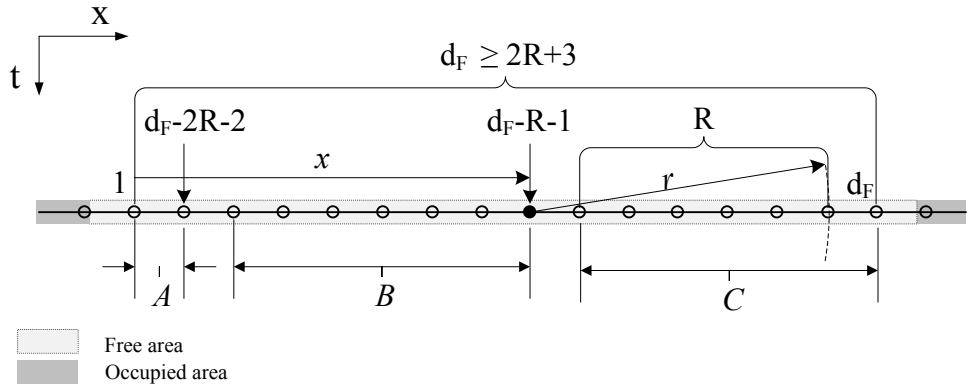

Figure B.5:   Example situation of calculating $p_V$ in case III

In area A, $1 \leq x \leq d_F - 2R - 2$, at the right side of $x$, $p_V(d_F, x)$ is the probability that no station starts transmit in the area from $x + 1$ to $x + 2R + 1$ in the next time slot. Therefore, we have

$$p_V(d_F, x) = (1 - p_{tx})^{2R+1}, 1 \leq x \leq d_F - 2R - 2 \tag{B.17}$$

In area $B$, $d_F - 2R - 1 \leq x \leq d_F - R - 1$, at the right side of $x$, $p_V(d_F, x)$ is the probability that no station starts transmit in the area from $x + 1$ to $d_F$ in the next time slot. Therefore,

$$p_V(d_F, x) = (1 - p_{tx})^{d_F - x}, d_F - 2R - 1 \leq x \leq d_F - R - 1 \tag{B.18}$$

In area $C$, $d_F - R \leq x \leq d_F$, any transmission will not cause a $V$ area at the right side of $x$. Therefore,

$$p_V(d_F, x) = 0, d_F - R \leq x \leq d_F \tag{B.19}$$

Bring Eq.(B.17), Eq.(B.18), and Eq.(B.19) into Eq.(B.11), we have

$$\overline{p'_{V|I}(d_F)} = \frac{1}{d_F} \sum_{x=1}^{d_F} R \cdot 2 \cdot p_V(d_F, x) \cdot p_{tx}(x)$$

$$= \frac{2R}{d_F} \cdot [(1 - p_{tx})^{R+1} + p_{tx} \cdot (1 - p_{tx})^{2R+1} \cdot (d_F - 2R - 2) - (1 - p_{tx})^{2R+2}]$$

$$, d_F \geq 2R + 3 \tag{B.20}$$

By calculating the mean value of $\overline{p'_{V|I}(d_F)}$ over all possible values of $d_F$, we get $p'_{V|I}$ in Eq.(B.21). Here, we use the geometric distribution of $d_F$ with parameter $p_{0,F}$.

$$p'_{V|I} = \frac{\sum_{n=1}^{\infty} \overline{p'_{V|I}(n)} \cdot n \cdot \Pr\{d_F = n | d_F \geq 1\}}{\sum_{n=1}^{\infty} n \cdot \Pr\{d_F = n | d_F \geq 1\}}$$

$$= \sum_{n=1}^{\infty} \overline{p'_{V|I}(n)} \cdot n \cdot (1 - p_{0,F})^{n-1} \cdot p_{0,F}^2$$

$$= 2 \cdot R \cdot p_{0,F}^2$$

$$\cdot \left\{ \frac{a^{R+1} \cdot [1 - (1 - p_{0,F})^{R+1}]}{p_{0,F}} - \frac{(1 - p_{tx}) \cdot a^{R+1} \cdot [1 - a^{R+1}]}{1 - a} \right.$$

$$+ \frac{[(1 - p_{tx})^{R+1} - [(2R + 1) \cdot p_{tx} + 1] \cdot (1 - p_{tx})^{2R+1}] \cdot (1 - p_{0,F})^{2R+2}}{p_{0,F}}$$

$$+ \left. \frac{[p_{tx} \cdot (1 - p_{tx})^{2R+1}] \cdot [(2R + 3) \cdot (1 - p_{0,F})^{2R+2} - (2R + 2) \cdot (1 - p_{0,F})^{2R+3}]}{p_{0,F}^2} \right\}$$

$$\tag{B.21}$$

## B.4 $p'_{B|I}$

$p'_{B|I}$ is the probability that at least one neighbor in the left- and in the right- channel sensing range of the station start to transmit in the next time slot simultaneously, conditioned on the station senses the channel idle at this time slot. Similar to the analysis of $p'_{I|I}$ in Section B.1, $p'_{B|I}$ is analyzed in three cases depending on the size of the channel free area $d_F$ at this time slot.

### B.4.1 Case I: $1 \leq d_F \leq R+1$

As shown in Figure B.1, in this case, $p'_{B|I}$ for the station at location $x$ is the probability that the station does not start transmit in the next time slot and at least one station in area $[1, x-1]$ and at least one station in area $[x+1, d_F]$ start transmit in the next time slot. Note: for $x = 1$ and $x = d_F$, $p'_{B|I} = 0$. Therefore, we have

$$p'_{B|I}(d_F, x) = (1 - p_{tx}) \cdot [1 - (1 - p_{tx})^{x-1}] \cdot [1 - (1 - p_{tx})^{d_F - x}]$$

$$, 1 \leq d_F \leq R + 1 \quad (B.22)$$

The mean value of $p'_{B|I}$ for a given $d_F$ in this case is

$$\overline{p'_{B|I}(d_F)} = \frac{1}{d_F} \sum_{x=1}^{d_F} p'_{B|I}(d_F, x)$$

$$= \frac{1}{d_F} \cdot (1 - p_{tx}) \cdot \left[ [1 + (1 - p_{tx})^{d_F - 1}] \cdot d_F - 2 \cdot \frac{1 - (1 - p_{tx})^{d_F}}{p_{tx}} \right]$$

$$, 1 \leq d_F \leq R + 1 \quad (B.23)$$

### B.4.2 Case II: $R+2 \leq d_F \leq 2R$

Three areas, namely $A$ for $1 \leq x \leq d_F - R - 1$, $B$ for $d_F - R \leq x \leq R + 1$, and $C$ for $R + 2 \leq x \leq d_F$, are identified in this case, as shown in Figure B.2.

For station at location $x$, $p'_{B|I}(d_F, x)$ is calculated for each area that $x$ falls in. Eq.(B.24) summarizes the results.

$$p'_{B|I}(d_F, x) = \begin{cases} (1 - p_{tx}) \cdot [1 - (1 - p_{tx})^{x-1}] \cdot [1 - (1 - p_{tx})^R] & , 1 \leq x \leq d_F - R - 1 \\ (1 - p_{tx}) \cdot [1 - (1 - p_{tx})^{x-1}] \cdot [1 - (1 - p_{tx})^{d_F - x}] & , d_F - R \leq x \leq R + 1 \\ (1 - p_{tx}) \cdot [1 - (1 - p_{tx})^R] \cdot [1 - (1 - p_{tx})^{d_F - x}] & , R + 2 \leq x \leq d_F \end{cases}$$

$$, R + 2 \leq d_F \leq 2R \quad (B.24)$$

The mean value of $p'_{B|I}(d_F, x)$ over all $x$ for a given $d_F$ in this case is

$$\overline{p'_{B|I}(d_F)} = \frac{1}{d_F}\sum_{x=1}^{d_F} p'_{B|I}(d_F, x)$$

$$= \frac{1}{d_F}\left\{2\cdot(1-p_{tx})\cdot[1-(1-p_{tx})^R]\cdot\left[(d_F-R-1) - \frac{1-(1-p_{tx})^{d_F-R-1}}{p_{tx}}\right]\right.$$

$$+ (1-p_{tx})$$

$$\left.\cdot\left[[1+(1-p_{tx})^{d_F-1}]\cdot(2R+2-d_F) - 2\cdot\frac{(1-p_{tx})^{d_F-R-1}-(1-p_{tx})^{R+1}}{p_{tx}}\right]\right\}$$

$$= \frac{1}{d_F}\cdot(1-p_{tx})$$

$$\cdot\left\{\left[\frac{4+2Rp_{tx}}{p_{tx}}\cdot(1-p_{tx})^R - \frac{2}{p_{tx}}\right] + \left[1-2(1-p_{tx})^R\right]d_F + \left[2R+2-\frac{2}{p_{tx}}\right]\right.$$

$$\left.\cdot(1-p_{tx})^{d_F-1} - d_F\cdot(1-p_{tx})^{d_F-1}\right\}$$

$$, R+2 \leq d_F \leq 2R \qquad (B.25)$$

### B.4.3 Case III: $d_F \geq 2R+1$

In this case, $p'_{B|I}(d_F, x)$ is calculated for $x$ values in the three areas identified in Figure B.3:

$$p'_{B|I}(d_F, x) = \begin{cases} (1-p_{tx})\cdot[1-(1-p_{tx})^{x-1}]\cdot[1-(1-p_{tx})^R] & , 1 \leq x \leq R \\ (1-p_{tx})\cdot[1-(1-p_{tx})^R]^2 & , R+1 \leq x \leq d_F-R \\ (1-p_{tx})\cdot[1-(1-p_{tx})^R]\cdot[1-(1-p_{tx})^{d_F-x}] & , d_F-R+1 \leq x \leq d_F \end{cases}$$

$$, d_F \geq 2R+1 \qquad (B.26)$$

The mean value of $p'_{B|I}(d_F, x)$ over all $x$ in this case is

$$\overline{p'_{B|I}(d_F)} = \frac{1}{d_F}\sum_{x=1}^{d_F} p'_{B|I}(d_F, x)$$

$$= \frac{1}{d_F}$$

$$\cdot\left\{2\cdot(1-p_{tx})\cdot[1-(1-p_{tx})^R]\cdot\left[R - \frac{1-(1-p_{tx})^R}{p_{tx}}\right] + (1-p_{tx})\right.$$

$$\left.\cdot[1-(1-p_{tx})^R]^2\cdot(d_F-2R)\right\}$$

$$, d_F \geq 2R+1 \qquad (B.27)$$

Taking the mean value of $\overline{p'_{B|I}(d_F)}$ over all possible $d_F$, we get $p'_{B|I}$ for steady state analysis. Here we use the observed geometric distribution of $d_F$ with parameter $p_{O,F}$ in this scenario.

$$p'_{B|I} = \frac{\sum_{n=1}^{\infty}\sum_{x=1}^{n} p'_{B|I}(n,x) \cdot \Pr\{d_F = n | d_F \geq 1\}}{\sum_{n=1}^{\infty} n \cdot \Pr\{d_F = n | d_F \geq 1\}}$$

$$= \sum_{n=1}^{\infty} \overline{p'_{B|I}(n)} \cdot n \cdot (1 - p_{O,F})^{n-1} \cdot p_{O,F}^2 \tag{B.28}$$

To conclude all three cases:

From Eq.(B.23) we have for $1 \leq d_F \leq R+1$

$$\sum_{n=1}^{R+1} \overline{p'_{B|I}(n)} \cdot n \cdot (1 - p_{O,F})^{n-1} \cdot p_{O,F}^2$$

$$= (1 - p_{tx}) \cdot p_{O,F}^2$$

$$\cdot \left[ -\frac{2}{p_{tx}} \sum_{n=1}^{R+1} (1 - p_{O,F})^{n-1} + \sum_{n=1}^{R+1} n \cdot (1 - p_{O,F})^{n-1} + \frac{2(1 - p_{tx})}{p_{tx}} \sum_{n=1}^{R+1} a^{n-1} \right.$$

$$\left. + \sum_{n=1}^{R+1} n a^{n-1} \right]$$

$$\tag{B.29}$$

Where, $\sum_{n=1}^{R+1} x^{n-1} = \frac{1-x^{R+1}}{1-x}$, $\sum_{n=1}^{R+1} nx^{n-1} = \frac{1-[(R+2)x^{R+1}-(R+1)x^{R+2}]}{(1-x)^2}$, and $a$ is given in Eq. (B.10).

From Eq.(B.25) we have for $R+2 \leq d_F \leq 2R$:

$$\sum_{n=R+2}^{2R} \overline{p'_{B|I}(n)} \cdot n \cdot (1 - p_{O,F})^{n-1} \cdot p_{O,F}^2$$

$$= (1 - p_{tx}) \cdot p_{O,F}^2$$

$$\cdot \left\{ \left[ \frac{4 + 2Rp_{tx}}{p_{tx}} \cdot (1 - p_{tx})^R - \frac{2}{p_{tx}} \right] \cdot \sum_{n=R+2}^{2R} (1 - p_{O,F})^{n-1} + [1 - 2(1 - p_{tx})^R] \right.$$

$$\cdot \sum_{n=R+2}^{2R} n \cdot (1 - p_{O,F})^{n-1} + \left[ 2R + 2 - \frac{2}{p_{tx}} \right] \cdot \sum_{n=R+2}^{2R} a^{n-1}$$

$$\left. - \sum_{n=R+2}^{2R} n \cdot a^{n-1} \right\}$$

$$\tag{B.30}$$

Where, $\sum_{n=R+2}^{2R} x^{n-1} = \frac{x^{R+1}-x^{2R}}{1-x}$, $\sum_{n=R+2}^{2R} nx^{n-1} = \frac{[(R+2)\cdot x^{R+1}-(R+1)\cdot x^{R+2}]-[(2R+1)\cdot x^{2R}-2R\cdot x^{2R+1}]}{(1-x)^2}$, and $a$ is given in Eq (B.10).

From Eq.(B.27) we have for $d_F \geq 2R + 1$:

$$\sum_{n=2R+1}^{\infty} \overrightarrow{p_{B|I}(n)} \cdot n \cdot (1 - p_{O,F})^{n-1} \cdot p_{O,F}^2 =$$

$$\left[ 2 \cdot (1 - p_{tx}) \cdot [1 - (1 - p_{tx})^R] \cdot \left[ R - \frac{1 - (1 - p_{tx})^R}{p_{tx}} \right] - (1 - p_{tx}) \cdot \left[ 1 - (1 - p_{tx})^R \right]^2 \cdot 2R \right]$$

$$\cdot p_{O,F} \cdot (1 - p_{O,F})^{2R} + (1 - p_{tx}) \cdot [1 - (1 - p_{tx})^R]^2$$

$$\cdot \left[ (2R + 1) \cdot (1 - p_{O,F})^{2R} - 2R \cdot (1 - p_{O,F})^{2R+1} \right]$$

(B.31)

## B.5 $p'_{VBL|I}$

According to the definition of $VBL$ state in Section 5.2.1[11], stations that can transit from state $I$ to state $VBL$ must be located between position 1 and $R$ in a free area, as shown in Figure B.6. $x - 1$ stations fall into the $VBL$ area if the left-most transmitter in the free area is at location $x$, $1 \leq x \leq R + 1$, in the next time slot. Note, here we only investigate the $VBL$ area at left side of the transmitter. Due to the symmetric property of the scenario, analysis of $VBL$ area of at the right side of the transmitter, i.e. for the free area between $d_F - R$ and $d_F$, yields the same results.

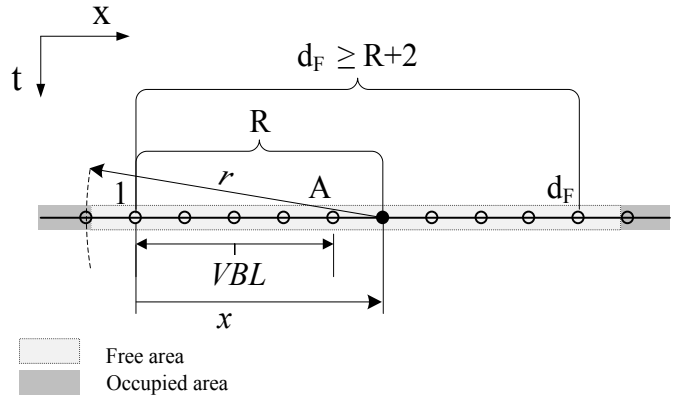

Figure B.6: Example situation of $VBL$ in case II $d_F \geq R+2$

As for $p'_{V|I}$, we use the ergodicity property of this scenario to calculate the mean probability $p'_{VBL|I}$. That is the probability each station transits from state $I$ to state $VBL$ equals to the mean percentage of stations transit from $I$ to $VBL$ in a free area of size $d_F$.

---

[11] Section 5.2.1 of [23] corresponds to Section VI.A.1 in the manuscript.

B.5.1 Case I: *1≤d_F≤R+1*

As shown in Figure B.7, the left-most transmitter at any location $x$ in this free area will cause $x - 1$ stations at the left side of the transmitter transit into $VBL$ state with probability $p'_{VBL|I}(d_F, x)$ in the next time slot. $p'_{VBL|I}(d_F, x)$ is calculated as the probability that the left-most transmitter is located at $x$ conditioned on the station at location $x$ starts to transmit in the next time slot:

$$p'_{VBL|I}(d_F, x) = (1 - p_{tx})^{(x-1)}, 1 \le d_F \le R + 1 \tag{B.32}$$

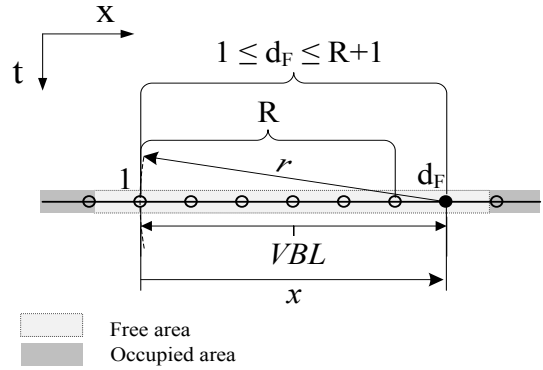

Figure B.7:   Example situation of *VBL* in case I *1≤d_F≤R+1*

The mean percentage of stations transit from $I$ to $VBL$ caused by the transmitter at location $x$ probability is

$$\overline{p'_{VBL|I}(d_F)} = 2 \cdot \sum_{x=1}^{d_F} \frac{(x-1)}{d_F}(1 - p_{tx})^{x-1} p_{tx}$$

$$= \frac{2(1 - p_{tx})}{d_F \cdot p_{tx}} \cdot \{1 - [d_F \cdot (1 - p_{tx})^{d_F - 1} - (d_F - 1) \cdot (1 - p_{tx})^{d_F}]\}$$

$$, 1 \le d_F \le R + 1 \tag{B.33}$$

Where, the coefficient 2 is multiplied because the calculation has to be done for the left side as well as the right side of $x$, which is symmetric in the scenario.

B.5.2 Case II: *d_F≥R+2*

As shown in Figure B.6, the only difference between case II and case I is that in case II only transmitters located at $x$, $1 \le x \le R + 1$, cause $x - 1$ stations at the left side of $x$ transiting from $I$ to $VBL$. Symmetrically, only transmitters at $x$, $d_F - R \le x \le d_F$, cause $d_F - x$ stations at the right side of $x$ transiting from $I$ to $VBL$. Therefore, in this case, the mean probability that a station in the free area transits from $I$ to $VBL$ in the next time slot for a given $d_F$ is

$$\overline{p'_{VBL|I}}(d_F) = 2 \cdot \sum_{x=1}^{R+1} \frac{(x-1)}{d_F}(1-p_{tx})^{x-1}p_{tx}$$

$$= \frac{2(1-p_{tx})}{d_F \cdot p_{tx}} \cdot \{1 - [(R+1) \cdot (1-p_{tx})^R - R \cdot (1-p_{tx})^{R+1}]\}$$

$$, d_F \geq R+1 \tag{B.34}$$

By calculating the mean value of $\overline{p'_{VBL|I}}(d_F)$ over all possible values of $d_F$, we get

$$p'_{VBL|I} = \frac{\sum_{n=1}^{\infty} \overline{p'_{VBL|I}}(n) \cdot n \cdot \Pr\{d_F = n | d_F \geq 1\}}{\sum_{n=1}^{\infty} n \cdot \Pr\{d_F = n | d_F \geq 1\}}$$

$$= \sum_{n=1}^{\infty} \overline{p'_{VBL|I}}(n) \cdot n \cdot (1-p_{O,F})^{n-1} \cdot p_{O,F}^2$$

$$= \frac{2 \cdot (1-p_{tx})}{p_{tx}} \cdot p_{O,F}^2$$

$$\cdot \left[ \frac{1 - (R+1) \cdot (1-p_{O,F}) \cdot a^R + R \cdot a^{R+1}}{p_{O,F}} \right.$$

$$- \frac{1 - (R+2) \cdot a^{R+1} + (R+1) \cdot a^{R+2}}{(1-a)^2} + (1-p_{tx}) \cdot a$$

$$\left. \cdot \frac{1 - (R+1) \cdot a^R + R \cdot a^{R+1}}{(1-a)^2} \right]$$

$$\tag{B.35}$$

Where, $a$ is given in Eq.(B.10).

Note: Eq.(B.35) uses the geometric distribution of $d_F$ with parameter $p_{O,F}$ for this scenario.

## B.6 $p'_{VBE|I}$

According to the time-domain Markov chain shown in Figure 5.4[12], the last supporting probability $p'_{VBE|I}$, when a station transits from state $I$, can be calculated from other supporting probabilities:

$$p'_{VBE|I} = 1 - p'_{I|I} - p'_{TX|I} - p'_{B|I} - p'_{V|I} - p'_{VBL|I} \tag{B.36}$$

---

[12] Figure 5.4 in [23] is numbered as Figure 6 in the manuscript.

## B.7 $p'_{B|V}$

According to the definition in Section 5.2.2.2[13], $R + 1 - x$ stations out of the $R$ stations in a hidden-station vulnerable area fall into the blocked area $B$ in the next time slot, if the closest later transmitter is at $x$, $1 \leq x \leq R$, in the adjacent channel free area, as shown in Figure B.8.

Similar to the analysis of $\overline{p'_{VBL|I}(d_F)}$ the mean percentage of stations in a hidden-station vulnerable area fall into area $B$ caused by a later transmission at location $x$ in the adjacent channel free area with size $d_F$ is calculated as:

$$\overline{p'_{B|V}(d_F)} = \sum_{x=1}^{d_F} \frac{(R+1-x)}{R} \cdot (1-p_{tx})^{x-1} \cdot p_{tx} \ , 1 \leq d_F \leq R \qquad (B.37)$$

And

$$\overline{p'_{B|V}(d_F)} = \sum_{x=1}^{R} \frac{(R+1-x)}{R} \cdot (1-p_{tx})^{x-1} \cdot p_{tx} \ , d_F \geq R+1 \qquad (B.38)$$

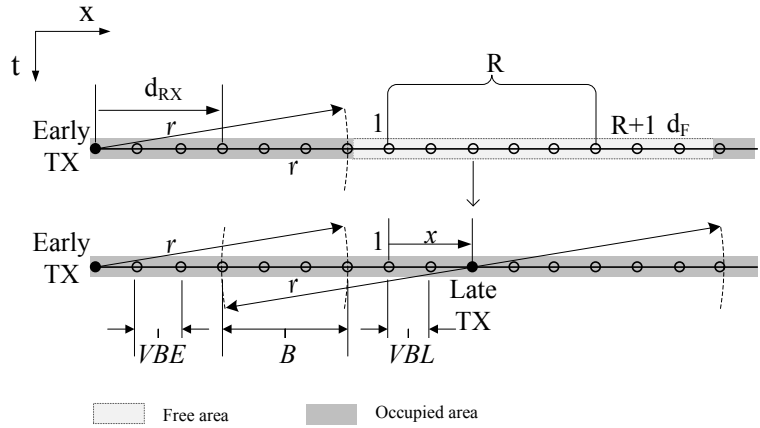

Figure B.8: Example situation of calculating $p'_{B|V}$

Following the same approach in the analysis of $p'_{VBL|I}$, we have

$$p'_{B|V} = \frac{\sum_{n=1}^{\infty} \overline{p'_{B|V}(n)} \cdot R \cdot \Pr\{d_F = n | d_F \geq 1\}}{\sum_{n=1}^{\infty} R \cdot \Pr\{d_F = n | d_F \geq 1\}}$$

$$= \sum_{n=1}^{\infty} \overline{p'_{B|V}(n)} \cdot (1-p_{O,F})^{n-1} \cdot p_{O,F}$$

---

[13] Section 5.2.2.2 of [23] corresponds to Section VI.A.1 in the manuscript.

$$= \frac{p_{tx} \cdot p_{O,F}}{R} \cdot \left[ \frac{(R+1) \cdot p_{tx} + (1-p_{tx}) \cdot a^R - 1}{p_{tx}^2 \cdot p_{O,F}} - \frac{(R+1) \cdot (1-p_{tx})}{p_{tx}} \cdot \frac{1-a^R}{1-a} \right.$$

$$+ \frac{1}{p_{tx}^2 \cdot (1-p_{O,F})} \cdot \frac{2a - a^2 - (R+2) \cdot a^{R+1} + (R+1) \cdot a^{R+2}}{(1-a)^2} - \frac{(1-p_{tx})^2}{p_{tx}^2}$$

$$\left. \cdot \frac{1 - (R+1) \cdot a^R + R \cdot a^{R+1}}{(1-a)^2} \right]$$

(B.39)

Where, $a$ is given in Eq.(B.10).

In addition to the mean transition probability $p'_{B|V}$, one can calculate the transition probability $p_{B|V}(d_{RX})$ at the receiver in state $V$ with topological distance $d_{RX}$ to the early transmitter, $1 \leq d_{RX} \leq R$, as shown in Figure B.8. Given the value of $d_F$, the conditional transition probability at $d_{RX}$ is calculated as:

$$\Pr\{V \to B \text{ at } d_{RX} | d_F = k\} = \begin{cases} 1 - (1-p_{tx})^{d_F} &, 1 \leq d_F \leq d_{RX} \\ 1 - (1-p_{tx})^{d_{RX}} &, d_F \geq d_{RX} + 1 \end{cases}$$

(B.40)

The mean transition probability from $V$ to $B$ at $d_{RX}$ is obtained by removing the condition on $d_F$ in Eq.(B.40):

$$p_{B|V}(d_{RX}) = \sum_{k=1}^{\infty} \Pr\{V \to B \text{ at } d_{RX} | d_F = k\} \cdot \Pr\{d_F = k\}$$

$$= \left[1 - (1-p_{O,F})^{d_{RX}}\right] - \frac{p_{O,F}}{1-p_{O,F}} \cdot \frac{a \cdot (1-a^{d_{RX}})}{1-a} + \left[1 - (1-p_{tx})^{d_{RX}}\right]$$

$$\cdot \left[(1-p_{O,F})^{d_{RX}}\right]$$

$$, 1 \leq d_{RX} \leq R \quad \text{(B.41)}$$

Where, the calculation is based on the geometric distribution of $d_F$ with the parameter $p_{O,F}$, and $a$ is given in Eq.(B.10).

## B.8 $p'_{VBE|V}$

The difference between the analyses of $p'_{B|V}$ and $p'_{VBE|V}$ is that instead of stations falling into area $B$, $p'_{VBE|V}$ is for stations falling into area $VBE$, as shown in Figure B.8. Besides, the later transmitter at $x$, $1 \leq x \leq R+1$, causes $x - 1$ stations in the hidden-station vulnerable area falling into the $VBE$.

Following the same approach in the analysis of $p'_{B|V}$, we have

$$\overline{p'_{VBE|V}}(d_F) = \sum_{x=1}^{d_F} \frac{(x-1)}{R} \cdot (1-p_{tx})^{x-1} \cdot p_{tx}$$

$$= \frac{(1-p_{tx})}{R \cdot p_{tx}} \cdot [1 - d_F \cdot (1-p_{tx})^{d_F-1} + (d_F - 1) \cdot (1-p_{tx})^{d_F}]$$

$$, 1 \le d_F \le R \tag{B.42}$$

And

$$\overline{p'_{VBE|V}}(d_F) = \sum_{x=1}^{R+1} \frac{(x-1)}{R} \cdot (1-p_{tx})^{x-1} \cdot p_{tx}$$

$$= \frac{(1-p_{tx})}{R \cdot p_{tx}} \cdot [1 - (R+1) \cdot (1-p_{tx})^R + R \cdot (1-p_{tx})^{R+1}]$$

$$, d_F \ge R + 1 \tag{B.43}$$

Consequently, the probability $p'_{VBE|V}$ is calculated as

$$p'_{VBE|V} = \frac{\sum_{n=1}^{\infty} \overline{p'_{VBE|V}}(n) \cdot R \cdot \Pr\{d_F = n | d_F \ge 1\}}{\sum_{n=1}^{\infty} R \cdot \Pr\{d_F = n | d_F \ge 1\}}$$

$$= \sum_{n=1}^{\infty} \overline{p'_{VBE|V}}(n) \cdot (1 - p_{O,F})^{n-1} \cdot p_{O,F}$$

$$= \frac{(1-p_{tx}) \cdot p_{O,F}}{R \cdot p_{tx}}$$

$$\cdot \left[ \frac{1 - a^R \cdot (1 + R \cdot p_{tx})}{p_{O,F}} - \frac{1 - (R+1) \cdot a^R + R \cdot a^{R+1}}{(1-a)^2} + (1-p_{tx}) \cdot a \right.$$

$$\left. \cdot \frac{1 - R \cdot a^{R-1} + (R-1) \cdot a^R}{(1-a)^2} \right]$$

$$\tag{B.44}$$

Where, $a$ is given in Eq.(B.10).

For a given $d_F$ value, the conditional transition probability from state $V$ to state $VBE$ at the receiver with topological distance $d_{RX}$, $1 \le d_{RX} \le R$, to the early transmitter, as shown in Figure B.8, is calculated as:

$$\Pr\{V \to VBE \text{ at } d_{RX} | d_F = k\}$$

$$= \begin{cases} 0 & , d_F \leq d_{RX} \\ (1 - p_{tx})^{d_{RX}} \cdot [1 - (1 - p_{tx})^{d_F - d_{RX}}] & , d_{RX} + 1 \leq d_F \leq R + 1 \\ (1 - p_{tx})^{d_{RX}} \cdot [1 - (1 - p_{tx})^{R+1-d_{RX}}] & , d_F \geq R + 2 \end{cases}$$

(B.45)

The mean transition probability from $V$ to $VBE$ at $d_{RX}$ is obtained by removing the condition on $d_F$ in Eq.(B.45):

$$p_{VBE|V}(d_{RX}) = \sum_{k=1}^{\infty} \Pr\{V \to VBE \text{ at } d_{RX} | d_F = k\} \cdot \Pr\{d_F = k\}$$

$$= (1 - p_{tx})^{d_{RX}} \cdot p_{O,F}$$

$$\cdot \left[ \frac{(1 - p_{O,F})^{d_{RX}} - (1 - p_{O,F})^{R+1}}{p_{O,F}} - \frac{a^{d_{RX}+1} \cdot (1 - a^{R-d_{RX}+1})}{(1 - p_{tx})^{d_{RX}} \cdot (1 - p_{O,F}) \cdot (1 - a)} \right]$$

$$+ (1 - p_{tx})^{d_{RX}} \cdot [1 - (1 - p_{tx})^{R-d_{RX}+1}] \cdot (1 - p_{O,F})^{R+1}$$

$$, 1 \leq d_{RX} \leq R \quad \text{(B.46)}$$

Where, the calculation is based on the geometric distribution of $d_F$ with the parameter $p_{O,F}$, and $a$ is given in Eq.(B.10).

## B.9 $p'_{V|V}$

$p'_{V|V}$ is identical for all stations in the hidden-station vulnerable area. It is the probability that no station between 1 and $R + 1$, as shown in Figure B.8, starts transmit in the next time slot. In other words, position 1 in the channel free area keeps free in the next time slot.

### B.9.1 Case I: $1 \leq d_F \leq R+1$

In this case, for all stations in the hidden-station vulnerable area,

$$\overline{p'_{V|V}(d_F)} = (1 - p_{tx})^{d_F} , 1 \leq d_F \leq R + 1 \quad \text{(B.47)}$$

### B.9.2 Case II: $d_F \geq R+2$

In this case, for all stations in the hidden-station vulnerable area,

$$\overline{p'_{V|V}(d_F)} = (1 - p_{tx})^{R+1} , d_F \geq R + 2 \quad \text{(B.48)}$$

$p'_{V|V}$ is calculated by taking the mean value of $\overline{p'_{V|V}(d_F)}$ over all possible values of $d_F$, given $d_F$ follows the geometric distribution with parameter $p_{O,F}$:

$$p'_{V|V} = \sum_{n=1}^{\infty} \overline{p'_{V|V}(d_F)} \cdot \Pr\{d_F = n | d_F \geq 1\} = \sum_{n=1}^{\infty} \overline{p'_{V|V}(d_F)} \cdot (1 - p_{O,F})^{n-1} \cdot p_{O,F}$$

$$= p_{O,F} \cdot (1 - p_{tx}) \cdot \frac{1 - a^{R+1}}{1 - a} + a^{R+1} \qquad (B.49)$$

Where, $a$ is given in Eq.(B.10).

Using Eq.(B.39), Eq.(B.44), and Eq.(B.49) we can prove:

$$p'_{B|V} + p'_{VBE|V} + p'_{V|V} = 1 \qquad (B.50)$$

For a given $d_F$ value, the conditional transition probability from state $V$ to state $V$ at the receiver with topological distance $d_{RX}$, $1 \leq d_{RX} \leq R$, to the early transmitter, as shown in Figure B.8, is calculated as:

$$\Pr\{V \to V \text{ at } d_{RX} | d_F = k\} = \begin{cases} (1 - p_{tx})^{d_F} & , 1 \leq d_F \leq R + 1 \\ (1 - p_{tx})^{R+1} & , d_F \geq R + 2 \end{cases}$$

$$(B.51)$$

The mean transition probability from $V$ to $V$ at $d_{RX}$ is obtained by removing the condition on $d_F$ in Eq.(B.51):

$$p_{V|V}(d_{RX}) = \sum_{k=1}^{\infty} \Pr\{V \to V \text{ at } d_{RX} | d_F = k\} \cdot \Pr\{d_F = k\} = \frac{p_{O,F}}{1 - p_{O,F}} \cdot \frac{a \cdot (1 - a^{R+1})}{1 - a} + a^{R+1}$$

$$, 1 \leq d_{RX} \leq R \qquad (B.52)$$

Where, the calculation is based on the geometric distribution of $d_F$ with the parameter $p_{O,F}$, and $a$ is given in Eq.(B.10).